\pdfoutput=1
\documentclass[prd,twocolumn,eqsecnum,nofootinbib,floatfix]{revtex4}
\usepackage{amssymb}
\usepackage{amsmath}
\usepackage{bm}
\usepackage{graphicx} 
\usepackage{float} 
\newcommand{\E}{{\cal E}} 
\newcommand{\B}{{\cal B}}

\newcommand{\e}[1]{{\text{e#1}}}
\begin{document}
\title{Relativistic theory of tidal Love numbers} 
\author{Taylor Binnington} 
\affiliation{Department of Physics, University of Guelph, Guelph,
  Ontario, N1G 2W1, Canada} 
\author{Eric Poisson} 
\affiliation{Department of Physics, University of Guelph, Guelph,
  Ontario, N1G 2W1, Canada;} 
\affiliation{Canadian Institute for Theoretical Astrophysics,
  University of Toronto, Toronto, Ontario, M5S 3H8, Canada} 
\date{September 16, 2009} 
\begin{abstract} 
In Newtonian gravitational theory, a tidal Love number relates the
mass multipole moment created by tidal forces on a spherical body to
the applied tidal field. The Love number is dimensionless, and it
encodes information about the body's internal structure. We present a
relativistic theory of Love numbers, which applies to compact bodies
with strong internal gravities; the theory extends and completes a
recent work by Flanagan and Hinderer, which revealed that the tidal
Love number of a neutron star can be measured by Earth-based
gravitational-wave detectors. We consider a spherical body deformed by
an external tidal field, and provide precise and meaningful
definitions for electric-type and magnetic-type Love numbers; and
these are computed for polytropic equations of state. The theory
applies to black holes as well, and we find that the relativistic Love
numbers of a nonrotating black hole are all zero. 
\end{abstract} 
\pacs{04.20.-q, 04.25.Nx, 04.40.Dg}
\maketitle

\section{Introduction and summary} 

\subsection*{Context of this work} 

The exciting prospect of using gravitational-wave detectors to measure
the tidal coupling of two neutron stars during the inspiral phase of
their orbital evolution was recently articulated by Flanagan and
Hinderer \cite{flanagan-hinderer:08, hinderer:08}. The idea is as
follows. The orbital motion of a binary
system of neutron stars produces the emission of gravitational waves,
which remove energy and angular momentum from the system. This causes 
the orbits to decrease in radius and increase in frequency, and leads
to the inspiraling motion of the compact bodies. Late in the inspiral
the gravitational waves enter the frequency band of the
detector, and detailed features of the orbital
motion are revealed in the shape and phasing of the wave. At the large
orbital separations that correspond to the low-frequency threshold of
the instrument, the tidal interaction between the bodies is
negligible, and the bodies behave as point masses. As the frequency
increases, however, the orbital separation decreases sufficiently that
the influence of the tidal interaction becomes important. The
bodies acquire a tidal deformation, and this affects their
gravitational field and orbital motion; the effect is revealed in the
shape and phasing of the gravitational waves. 

Flanagan and Hinderer have provided a quantitative analysis of this
story, and they have shown that the tidal coupling between neutron
stars is accessible to measurement by the current generation of
Earth-based gravitational-wave detectors (such as Enhanced
LIGO). This prospect is exciting, because the details of the tidal
interaction depend on the internal structure of each body, and the
measurement can thus reveal important information regarding the
compactness of each body, as well as its equation of state; and this
information is released cleanly, during the inspiral phase of the
orbital evolution, well before the messy merger of the two
companions.   

\subsection*{Newtonian theory of tidal Love numbers} 

The effect of the tidal interaction on the orbital motion and
gravitational-wave signal is measured by a quantity
known as the {\it tidal Love number} of each companion
\cite{love:11}. In Newtonian gravity (see, for example, 
Ref.~\cite{murray-dermott:99}), the tidal Love number is a constant of
proportionality between the tidal field applied to the body and the
resulting multipole moment of its mass distribution. In the
quadrupolar case, the tidal field is characterized by the 
{\it tidal moment} $\E_{ab}(t) := -\partial_{ab} U_{\rm ext}$, in 
which the external Newtonian potential $U_{\rm ext}$ is sourced by the 
companion body and evaluated (after differentiation with respect to
the spatial coordinates) at the body's center-of-mass. Because
the external potential satisfies Laplace's equation in the body's
neighborhood, the tidal-moment tensor is not only symmetric but also
tracefree; it is a symmetric-tracefree (STF) tensor. 

The quadrupole moment is $Q^{ab} := \int \rho (x^a x^b - \frac{1}{3}  
\delta^{ab} r^2)\, d^3x$, where $\rho$ is the mass density inside the  
body, $x^a$ is a Cartesian coordinate system whose origin is at the
center-of-mass, and $r := (\delta_{ab} x^a x^b)^{1/2}$ is the distance
to the center-of-mass; the quadrupole moment is another STF
tensor. In the absence of a tidal field the body would be spherical,
and its quadrupole moment would vanish. In the presence of a (weak)
tidal field, the quadrupole moment is proportional to the tidal field,
and dimensional analysis requires an expression of
the form $Q_{ab} = -\frac{2}{3} k_2 R^5 \E_{ab}$. (We use relativistic 
units and set $G = c =1$.) Here $R$ is the body's radius, and the
factor of $\frac{2}{3}$ is conventional; the dimensionless constant
$k_2$ is the tidal Love number for a quadrupolar deformation. Using
these expressions, the Newtonian potential outside the body can be
written as a sum of body and external potentials, and we have 
\begin{equation} 
U = \frac{M}{r} 
- \frac{1}{2} \bigl[ 1 + 2k_2 (R/r)^5 \bigr] \E_{ab}(t) x^a x^b.  
\label{Newt_l2} 
\end{equation} 
The first term is evidently the monopole piece of the potential, which
depends on the body's mass $M$. Within the square brackets, the first
term represents the applied tidal field, and the second term is the
body's response, measured in terms of the Love number $k_2$.   

In Eq.~(\ref{Newt_l2}) the total potential was truncated to the
leading, quadrupole order in a Taylor expansion of the external
potential; additional terms would involve tidal moments of higher 
multipole orders, and higher powers of the coordinates $x^a$.  When
the tidal field is a pure multipole of order $l$, Eq.~(\ref{Newt_l2})
generalizes to 
\begin{equation} 
U = \frac{M}{r} - \frac{1}{(l-1)l} \bigl[ 1 
+ 2k_l (R/r)^{2l+1} \bigr] \E_{L}(t) x^L. 
\label{Newt} 
\end{equation} 
Here $k_l$ is the Love number for this multipolar configuration, and
$L := a_1 a_2 \cdots a_l$ is a multi-index that contains a number $l$
of individual indices. The tidal moment is now defined by $\E_L(t) :=
-\partial_L U_{\rm ext}/(l-2)!$, and it is symmetric and tracefree in
all pairs of indices. We also introduced $x^L := x^{a_1} x^{a_2}
\cdots x^{a_l}$. In this generalized case the $l$-pole moment of the
mass distribution is the STF tensor $Q^L := \int \rho 
x^{\langle L \rangle}\, d^3x$, where the angular brackets indicate
that all traces must be removed from the tensor $x^L$; it is related
to the tidal moment by $Q_L = -[2(l-2)!/(2l-1)!!] k_l R^{2l+1} \E_L$.    

\subsection*{Purpose of this work} 

Our purpose in this paper is to introduce a precise notion of tidal Love
numbers in general relativity, something that was not pursued in the
original work by Flanagan and Hinderer \cite{flanagan-hinderer:08,
  hinderer:08}. In fact, we provide precise
definitions for two types of tidal Love numbers: an electric-type Love
number $k_{\rm el}$ that has a direct analogy with the Newtonian Love 
number introduced previously, and a magnetic-type Love number 
$k_{\rm mag}$ that has no analogue in Newtonian gravity. Magnetic-type  
Love numbers were introduced in post-Newtonian theory in the works of
Damour, Soffel, and Xu \cite{damour-soffel-xu:92} and Favata
\cite{favata:06}. Our definitions apply to gravitational fields that
are arbitrarily strong, and to (weak) tidal deformations of any
multipolar order.   

Our relativistic Love numbers are defined within the context of linear
perturbation theory, in which an initially-spherical body is perturbed
slightly by an applied tidal field. Our definitions are restricted to
slowly-changing tidal fields; this means that while a tidal moment
such as $\E_L(t)$ does depend on time, to reflect the changes in the
external distribution of matter, the dependence is sufficiently slow
that the body's response presents only a {\it parametric dependence}
upon time. This allows us to ignore time-derivative terms in the field
equations, because they are much smaller than the spatial-derivative
terms. For all practical purposes the perturbation is stationary, and
$t$ appears as an adiabatic parameter.  

Gravitational perturbations of spherically-symmetric bodies
are described by a metric perturbation $p_{\alpha\beta}$ that can be
decomposed into tensorial spherical harmonics; each multipole can be
considered separately. The complete spacetime metric is 
$g_{\alpha\beta} = g^0_{\alpha\beta} + p_{\alpha\beta}$, with 
$g^0_{\alpha\beta}$ denoting the (spherically-symmetric) metric of the
unperturbed body. We work in the body's immediate neighborhood, and
the external bodies that create the (multipolar) tidal field are
assumed to live outside this neighborhood. To {\it define} the
relativistic Love numbers it is sufficient to consider the vacuum
region external to the body, and to construct $g_{\alpha\beta}$ in
this region only; this metric will be a solution to the vacuum field
equations, and will represent the relativistic generalization
of Eq.~(\ref{Newt}). To {\it compute} the Love numbers it is necessary
to construct $g_{\alpha\beta}$ in the body's interior also, and this
requires the formulation of a stellar model. The external problem
therefore applies to any type of body, while the internal problem
refers to a specific choice of equation of state. 

\subsection*{External problem} 

We review the external problem first. We erect a coordinate system
$(v,r,\theta,\phi)$ that is intimately tied to the behavior of light
cones: The advanced-time coordinate $v$ is constant on past light
cones that converge toward the center at $r=0$, $r$ is both an areal
radius and an affine-parameter distance along the null generators of
each light cone, and the angular coordinates $\theta^A =
(\theta,\phi)$ are constant on each generator. This choice of
coordinates is inherited from previous work on the tidal
deformation of black holes \cite{poisson:05}. 

In these coordinates the external metric of the 
{\it unperturbed body} is given by $ds_0^2 = -f\, dv^2 
+ 2\, dvdr + r^2 d\Omega^2$, in which $f := 1-2M/r$ and 
$d\Omega^2 := d\theta^2 + \sin^2\theta\, d\phi^2$; this is 
the Schwarzschild metric presented in Eddington-Finkelstein
coordinates. To construct the perturbation we impose the 
{\it light-cone gauge conditions} $p_{vr} = p_{rr} = p_{r\theta} =
p_{r\phi} = 0$ to ensure that the coordinates keep their geometrical 
meaning in the perturbed spacetime \cite{preston-poisson:06b}. (This
property makes the light-cone gauge superior to the popular
Regge-Wheeler gauge, which does not provide the coordinates with any
geometrical meaning.) A perturbation of multipole order $l$ can be
decomposed into even-parity and odd-parity sectors, and each sector
must be a solution to the Einstein field equations linearized about the
Schwarzschild metric.  

The even-parity sector is generated by the electric-type tidal moment
$\E_L(v)$, an STF tensor defined in a quasi-Cartesian system $x^a$
related in the usual way to the spherical coordinates
$(r,\theta^A)$. The $(2l+1)$ independent components of this tensor can 
be encoded in the functions $\E^{(l)}_m(v)$, in which the azimuthal
index $m$ is an integer within the interval $-l \leq m \leq l$; the
encoding is described by $\E_L x^L = r^l \sum_m \E^{(l)}_m
Y^{lm}(\theta^A)$, in which $Y^{lm}$ are the usual spherical-harmonic 
functions. We define the tidal potentials  
\begin{subequations} \allowdisplaybreaks
\label{Epot}  
\begin{align} 
\E^{(l)}(v,\theta^A) &= \sum_m \E^{(l)}_m(v) Y^{lm}(\theta^A), \\ 
\E^{(l)}_A(v,\theta^A) &= \frac{1}{l} \sum_m \E^{(l)}_m(v) 
Y_A^{lm}(\theta^A), \\ 
\E^{(l)}_{AB}(v,\theta^A) &= \frac{2}{l(l-1)} \sum_m \E^{(l)}_m(v)
Y_{AB}^{lm}(\theta^A), 
\end{align}
\end{subequations} 
in which $Y^{lm}_A$ and $Y^{lm}_{AB}$ are vector and tensor spherical 
harmonics of even parity; these are defined in Sec.~II. 

The odd-parity sector is generated by the magnetic-type tidal moment
$\B_L(v)$, another STF tensor whose independent components can be
encoded (as previously) in the functions $\B^{(l)}_m(v)$. The
odd-parity tidal potentials are  
\begin{subequations} \allowdisplaybreaks
\label{Bpot}  
\begin{align} 
\B^{(l)}_A(v,\theta^A) &= \frac{1}{l} \sum_m \B^{(l)}_m(v) 
X_A^{lm}(\theta^A), \\ 
\B^{(l)}_{AB}(v,\theta^A) &= \frac{2}{l(l-1)} \sum_m \B^{(l)}_m(v)
X_{AB}^{lm}(\theta^A), 
\end{align}
\end{subequations} 
in which $X^{lm}_A$ and $X^{lm}_{AB}$ are vector and tensor spherical
harmonics of odd parity; these also are defined in Sec.~II. There is
no scalar potential $\B^{(l)}$ in the odd-parity sector. 

The metric outside any spherical body deformed by a tidal environment
characterized by the tidal moments $\E_L$ and $\B_L$ is calculated in
Sec.~III. It is given by 
\begin{subequations} \allowdisplaybreaks
\label{deformed_metric} 
\begin{align} 
g_{vv} &= -f -\frac{2}{(l-1)l} r^l e_1(r) \E^{(l)}, \\ 
g_{vr} &= 1, \\
g_{vA} &= -\frac{2}{(l-1)(l+1)} r^{l+1} e_4(r) \E^{(l)}_A 
\nonumber \\ & \qquad \mbox{} 
+ \frac{2}{3(l-1)} r^{l+1} b_4(r) \B^{(l)}_A, \\ 
g_{AB} &= r^2 \Omega_{AB} 
- \frac{2}{l(l+1)} r^{l+2} e_7(r) \E^{(l)}_{AB} 
\nonumber \\ & \qquad \mbox{} 
+ \frac{2}{3l} r^{l+2} b_7(r) \B^{(l)}_{AB}. 
\end{align} 
\end{subequations} 
The radial functions are 
\begin{subequations} \allowdisplaybreaks 
\label{radial} 
\begin{align} 
e_1 &=A_1 + 2 k_{\rm el} (R/r)^{2l+1} B_1, \\ 
e_4 &= A_4 - 2\frac{l+1}{l} k_{\rm el} (R/r)^{2l+1} B_4, \\ 
e_7 &= A_7 + 2 k_{\rm el} (R/r)^{2l+1} B_7, \\ 
b_4 &= A_4 - 2\frac{l+1}{l} k_{\rm mag} (R/r)^{2l+1} B_4, \\ 
b_7 &= A_7 + 2 k_{\rm mag} (R/r)^{2l+1} B_7, 
\end{align} 
\end{subequations} 
with 
\begin{subequations} \allowdisplaybreaks 
\label{functions} 
\begin{align} 
A_1 &:= f^2 F(-l+2,-l;-2l;2M/r), \\ 
B_1 &:= f^2 F(l+1,l+3;2l+2;2M/r), \\ 
A_4 &:= F(-l+1,-l-2;-2l;2M/r), \\
B_4 &:= F(l-1,l+2;2l+2;2M/r), \\
A_7 &:= \frac{l+1}{l-1} F(-l,-l;-2l;2M/r) 
\nonumber \\ & \qquad \mbox{} 
- \frac{2}{l-1} F(-l,-l-1;-2l;2M/r), \\ 
B_7 &:= \frac{l}{l+2} F(l+1,l+1;2l+2;2M/r) 
\nonumber \\ & \qquad \mbox{} 
+ \frac{2}{l+2} F(l,l+1;2l+2;2M/r). 
\end{align}
\end{subequations} 
Here $R$ is the body's radius, and $F(a,b;c;z)$ is the hypergeometric
function. The functions $A_n$ are finite polynomials in $2M/r$, while
the functions $B_n$ have non-terminating expansions in powers of
$2M/r$; for selected values of $l$ they can be expressed in terms of
elementary functions such as $\ln(1-2M/r)$ and finite polynomials (see
Table I in Sec.~III). Each one of these functions goes to one as $r$
goes to infinity. And while $A_n$ is finite at $r=2M$, we observe that
$B_n$ diverges logarithmically when $r \to 2M$.    

The metric of Eqs.~(\ref{deformed_metric}) is valid in a neighborhood 
of the deformed body, and it provides a definition for the
electric-type Love numbers $k_{\rm el}$ and the magnetic-type Love
numbers $k_{\rm mag}$; these refer to the multipole order $l$, but we
suppress the use of this label to keep the notation
clean. While the definitions seem to rely on a specific choice of
gauge for the metric perturbation, we prove in Sec.~III that our Love 
numbers are gauge-invariant.  

When the tidal moments are switched off the metric reduces to the
Schwarzschild metric expressed in the light-cone coordinates
$(v,r,\theta^A)$. When the mass parameter $M$ is set equal to zero the
metric describes the neighborhood of a geodesic world line in a
Ricci-flat spacetime. In this limit the tidal moments can be related
to the derivatives of the Weyl tensor evaluated at $r=0$. According to  
Eqn.~(1.3) of Ref.~\cite{zhang:86}, we have that $\E_L 
= [(l-2)!]^{-1} (C_{t a_1 t a_2 ; a_3 \cdots a_l})^{\rm STF}$ and
$\B_L = [\frac{2}{3}(l+1)(l-2)!]^{-1} ( \epsilon_{a_1 b c} 
C^{bc}_{\ \ a_2 t ; a_3 \cdots a_l} )^{\rm STF}$, where $\epsilon_{abc}$
is the permutation symbol and the tensor components are listed in the
quasi-Lorentzian coordinates $(t := v-r, x^a)$; the STF superscript
indicates that the $a_n$ indices are symmetrized and all traces are
removed. In the spacetime of Eq.~(\ref{deformed_metric}) the tidal
moments $\E_L$ and $\B_L$ retain a similar relationship with 
the Weyl tensor, with the understanding that the relations are now
approximate and refer to the asymptotic behavior of the Weyl tensor
for $r \gg M$.     

The perturbed metric of Eq.~(\ref{deformed_metric}) can be compared
with the Newtonian potential of Eq.~(\ref{Newt}). We define an
effective Newtonian potential $U_{\rm eff}$ by 
$g_{vv} =: -(1-2U_{\rm eff})$, and our expression for
$g_{vv}$ implies that in general relativity, 
\begin{equation} 
U_{\rm eff} = -\frac{M}{r} - \frac{1}{(l-1)l} \Bigl[
A_1 + 2 k_{\rm el} (R/r)^{2l+1} B_1 \Bigr] \E_L(v) x^L. 
\end{equation} 
In the nonrelativistic limit, $A_1$ and $B_1$ are both approximately
equal to unity, and we recover Eq.~(\ref{Newt}); the electric-type
Love number $k_{\rm el}$ reduces to the Newtonian number $k_l$. In the
strong-field regime we still recognize the $A_1$ term as coming from
the applied tidal field, while the $B_1$ term is clearly associated
with the body's response. There is no confusion between these terms,
because the structure of $A_1$ is that of the finite polynomial $1 +
\cdots + \lambda (2M/r)^l$, which does not contain a term of order
$(2M/r)^{2l+1}$; $\lambda$ is a numerical factor that can be
determined by expanding the hypergeometric function. Because $r$ is
geometrically well defined, we can always distinguish the tidal terms
from the body terms in the metric.  

The light-cone coordinates $(v,r,\theta^A)$ are well-behaved across an 
eventual event horizon of the perturbed spacetime, and our formalism
is capable of handling black holes as well as material bodies. In
general, however, the metric of Eqs.~(\ref{deformed_metric}) is not
regular at the event horizon, because of the presence of the $B_n$
functions, which diverge logarithmically in the limit $r \to 2M$. To
represent a perturbed black hole the metric must be devoid of these
terms, and this can be accomplished by assigning 
$k_{\rm el} = k_{\rm mag} = 0$ to a black hole. This is one of the
major conclusions of this work:  
{\it The relativistic Love numbers of a nonrotating black hole are all
zero.} This result is contained implicitly in Ref.~\cite{poisson:05}, 
but the formalism of this paper permits a much clearer articulation of 
this property.  

\subsection*{Internal problem} 

To {\it compute} the relativistic Love numbers for a selected stellar
model requires the construction of the internal metric (also expressed
as a sum of unperturbed solution and linear perturbation) and its
matching with the external metric at the perturbed boundary of the
matter distribution. We carry out this exercise in Secs.~IV and V,
adapting the formalism of Thorne and Campolattaro 
\cite{thorne-campolattaro:67} to our light-cone coordinates. We take
the body to consist of a perfect fluid with a polytropic equation of
state   
\begin{equation} 
p = K \rho^{1 + 1/n}. 
\end{equation} 
Here $p$ is the fluid's pressure, $\rho$ its proper energy density,
$K$ is a constant, and $n$ is the polytropic index (another constant).  

Our results are presented in Figs.~1--8, and tables of values are
provided in the Appendix. In each figure we plot the Love number
for a selected multipole order (from $l=2$ to $l=5$), and for selected 
values of the polytropic index $n$ (from $n=0.5$ to $n=2.0$), as a
function of the stellar compactness parameter $C := 2M/R$; this ranges
from $C=0$ --- a weak-field, Newtonian configuration --- to 
$C = C_{\rm max}$, with $C_{\rm max}$ representing the compactness of 
the maximum-mass configuration for the selected equation of state.  

For the electric-type Love numbers we observe the following
features. (i) At $C=0$ we recover the Newtonian values for polytropes,
as tabulated by Brooker and Olle \cite{brooker-olle:55}. (ii) For a
constant $C$, $k_{\rm el}$ decreases as the polytropic index
increases; this reflects the fact that as $n$ increases, the matter
distribution becomes increasingly concentrated near the center, which 
inhibits the development of large multipole moments. (iii) For a
constant $n$, $k_{\rm el}$ decreases as the compactness parameter
increases; this reflects the fact that as $C$ increases, the strength
of the internal gravity increases, which produces an increased
resistance to tidal deformations.     

For the magnetic-type Love numbers we observe the following
features. (i) At $C=0$ the Love numbers are all zero; this reflects
the fact that the magnetic-type tidal coupling is a purely
relativistic effect that has a vanishing Newtonian limit. (ii) For a
constant $C$, $k_{\rm mag}$ decreases as the polytropic index
increases; this is explained as in the preceding paragraph. (iii) For
a constant $n$, $k_{\rm mag}$ first increases as $C$ increases, but 
then it decreases after reaching a maximum; this reflects the fact
that the magnetic-type tidal coupling is the result of an internal
competition: a strong field is required to produce an effect in the
first place, but it eventually causes a large resistance to
tidal deformation. 

\begin{figure}
\includegraphics[width=0.9\linewidth]{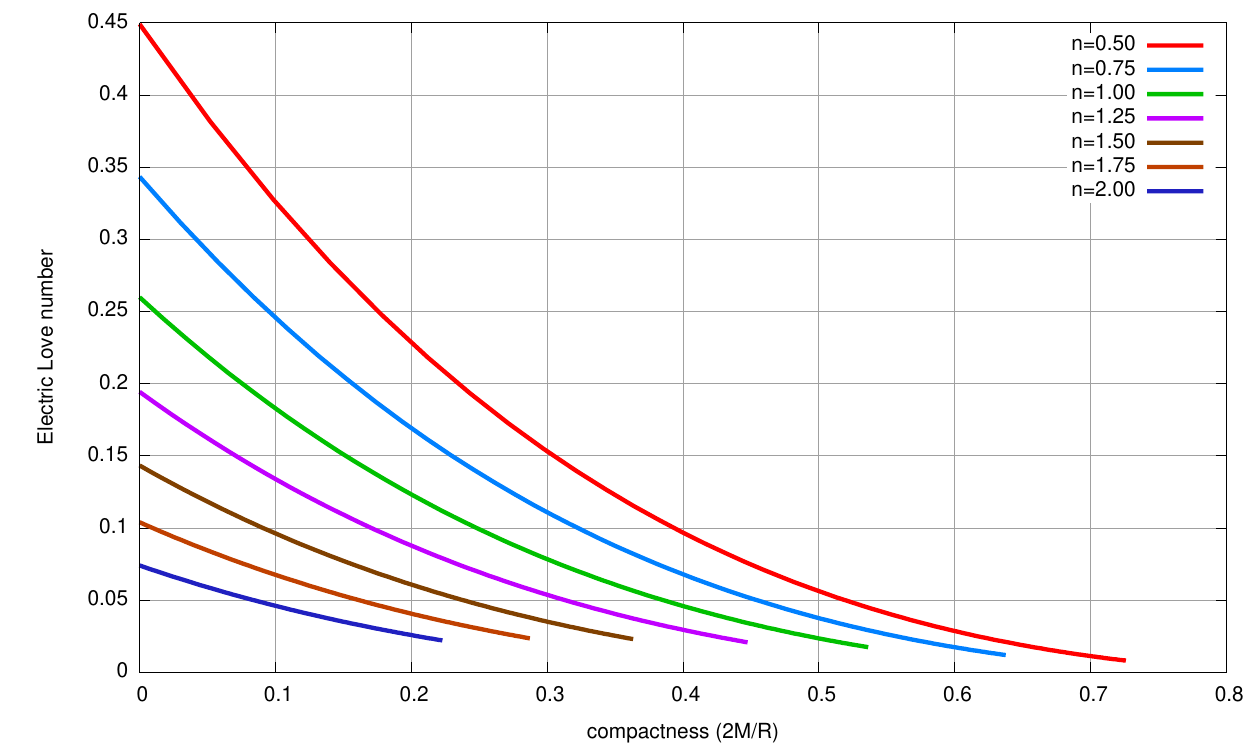}
\caption{Electric-type Love numbers for $l=2$, plotted as functions of
  the compactness parameter $2M/R$. The uppermost curve corresponds to
  $n=0.5$ and the stiffest equation of state. The lowermost curve
  corresponds to $n=2.0$ and the softest equation of state. The
  curves in between are ordered by the value of $n$. The arrangement
  is the same in all other figures.} 
\end{figure}

\begin{figure}
\includegraphics[width=0.9\linewidth]{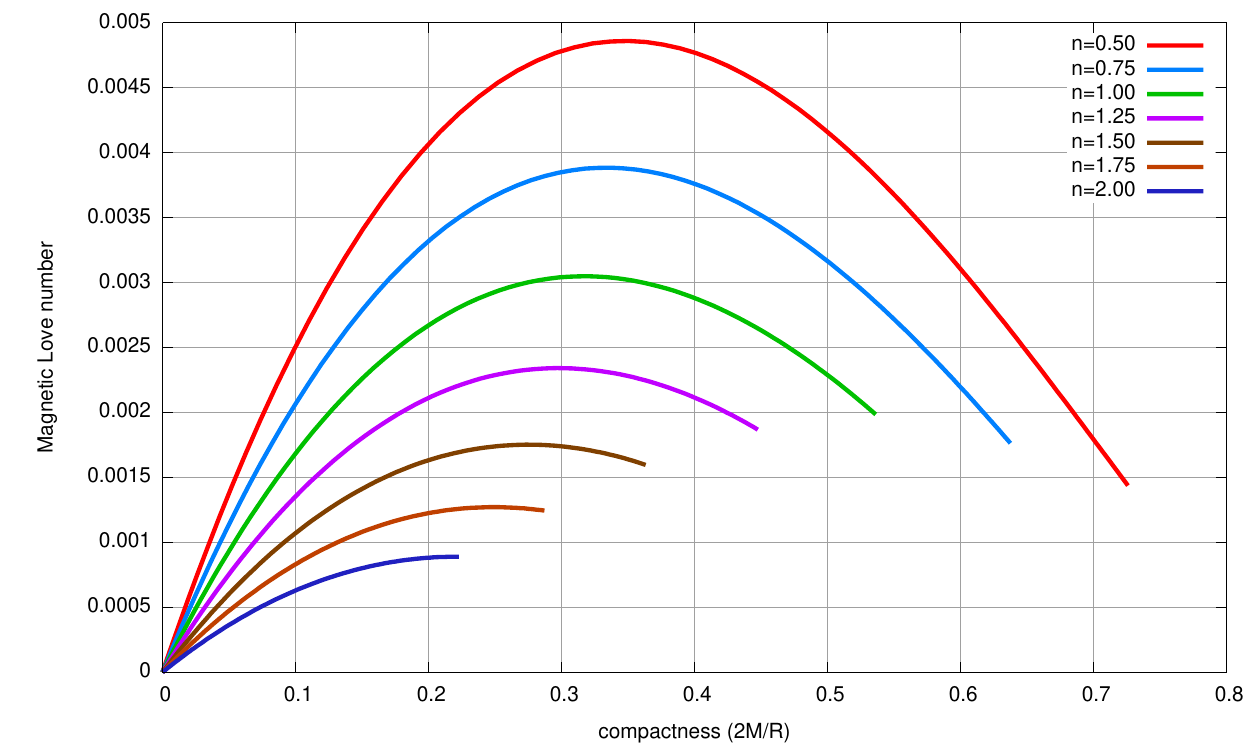}
\caption{Magnetic-type Love numbers for $l=2$, plotted as functions of
  the compactness parameter $2M/R$.} 
\end{figure}

\begin{figure}
\includegraphics[width=0.9\linewidth]{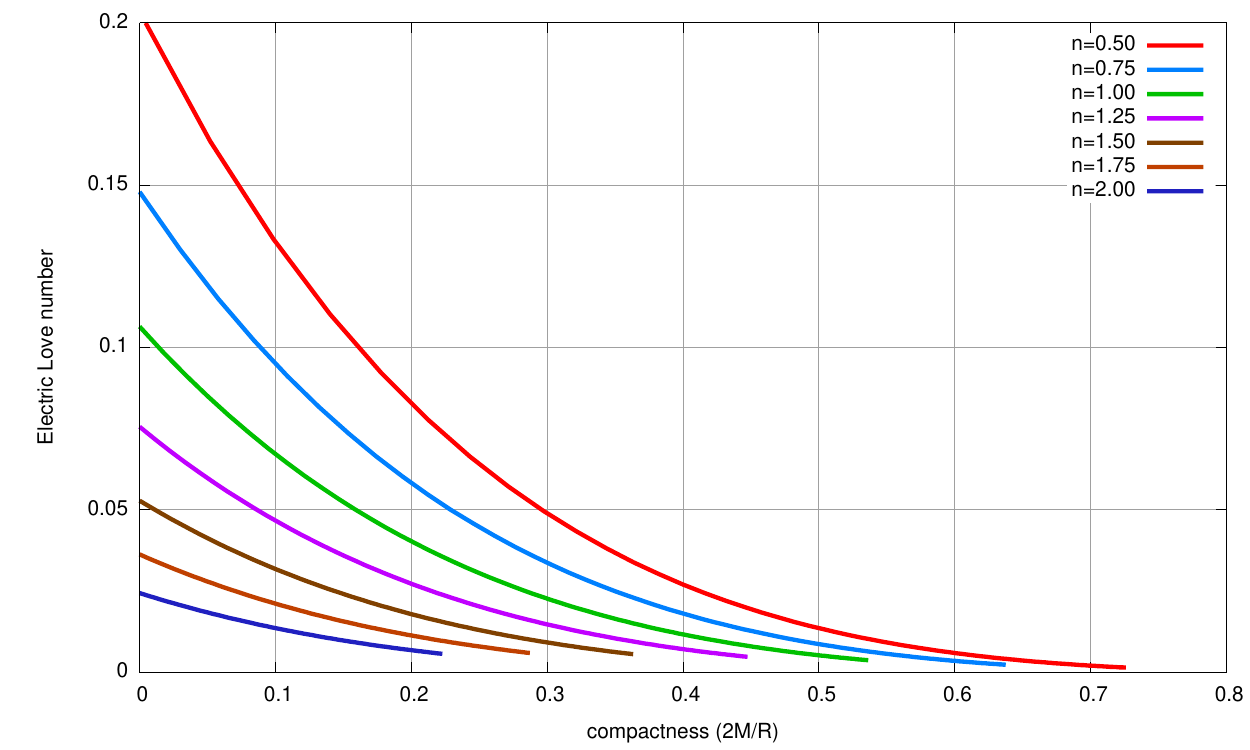}
\caption{Electric-type Love numbers for $l=3$.} 
\end{figure}

\begin{figure}
\includegraphics[width=0.9\linewidth]{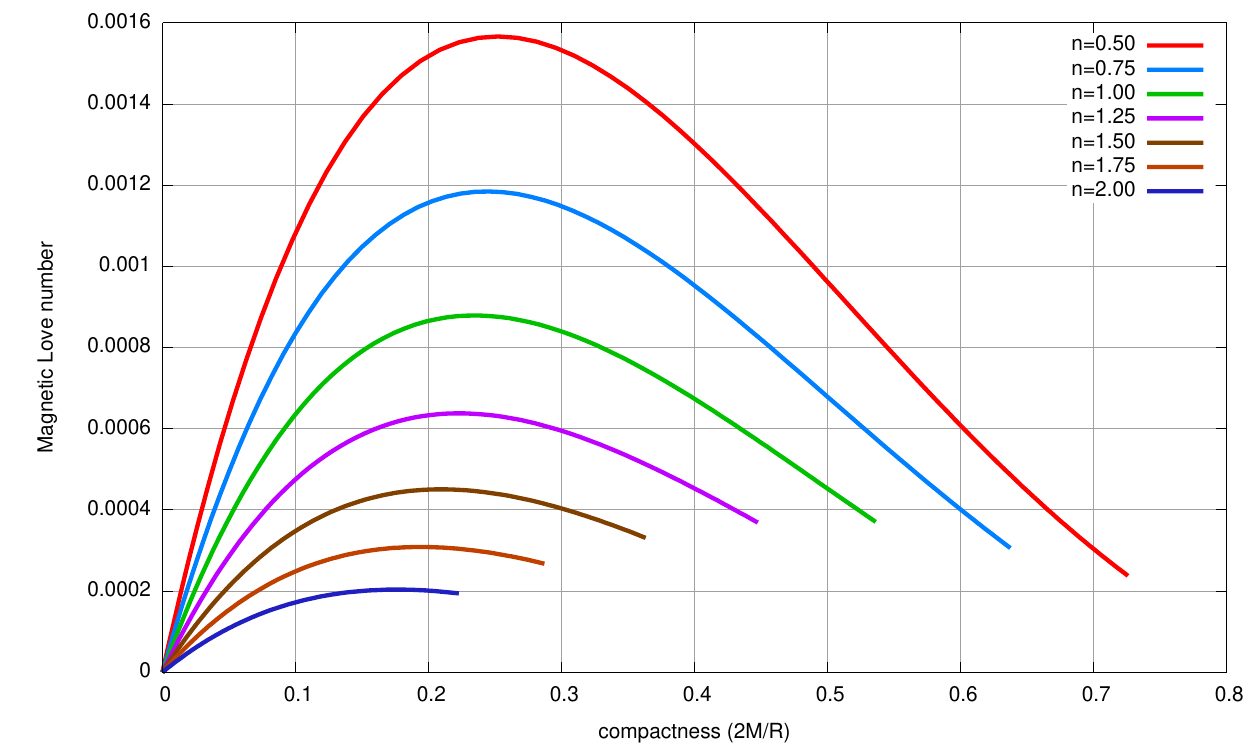}
\caption{Magnetic-type Love numbers for $l=3$.} 
\end{figure}

\begin{figure}
\includegraphics[width=0.9\linewidth]{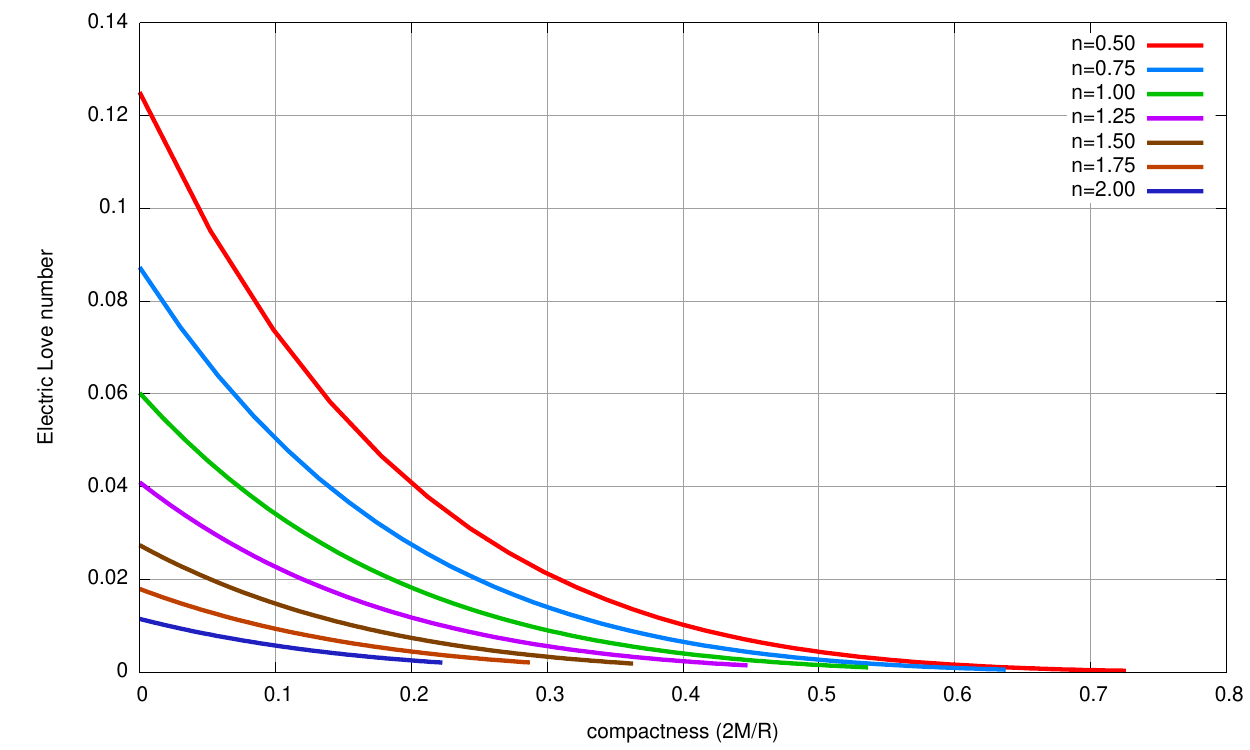}
\caption{Electric-type Love numbers for $l=4$.} 
\end{figure}

\begin{figure}
\includegraphics[width=0.9\linewidth]{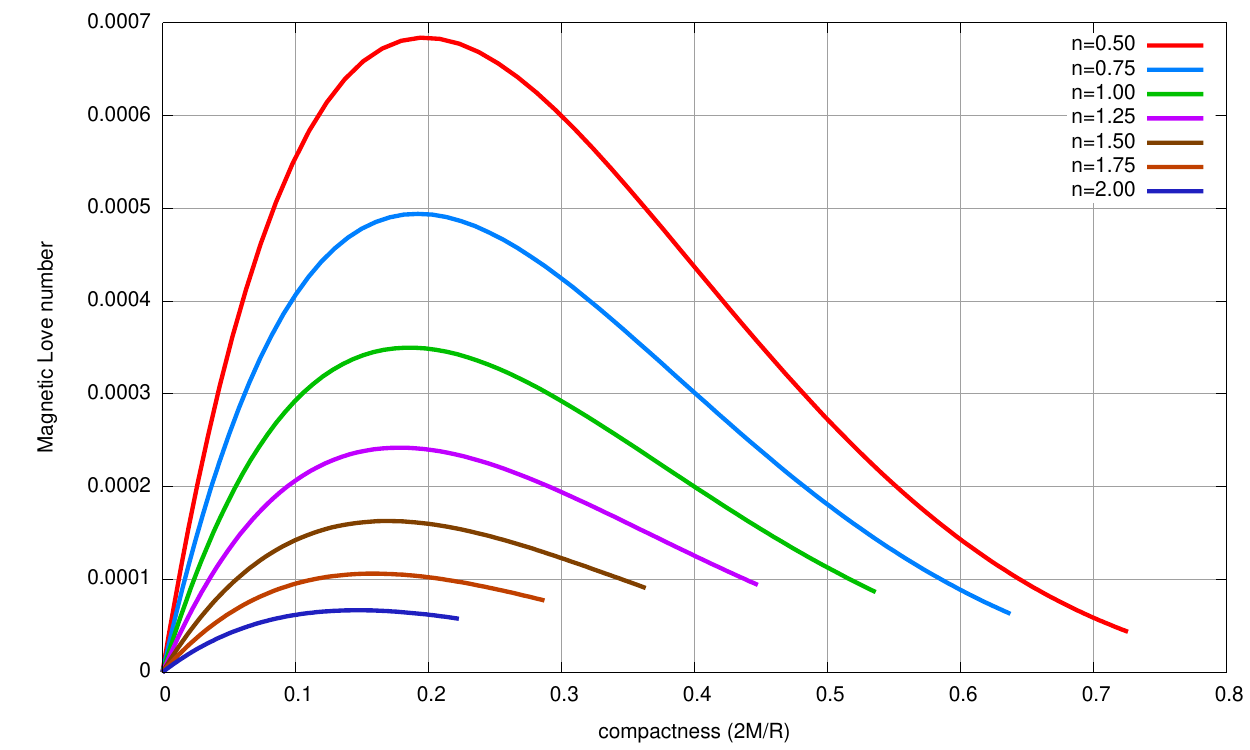}
\caption{Magnetic-type Love numbers for $l=4$.} 
\end{figure}

\begin{figure}
\includegraphics[width=0.9\linewidth]{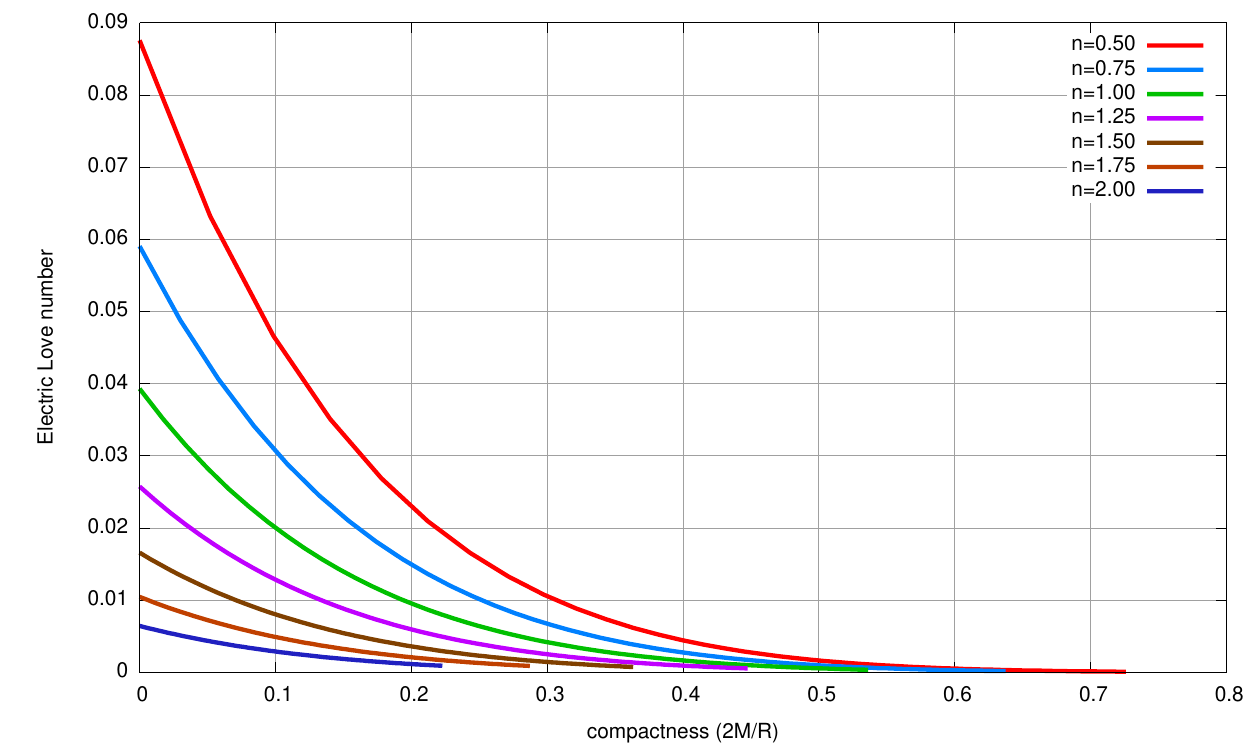}
\caption{Electric-type Love numbers for $l=5$.} 
\end{figure}

\begin{figure}
\includegraphics[width=0.9\linewidth]{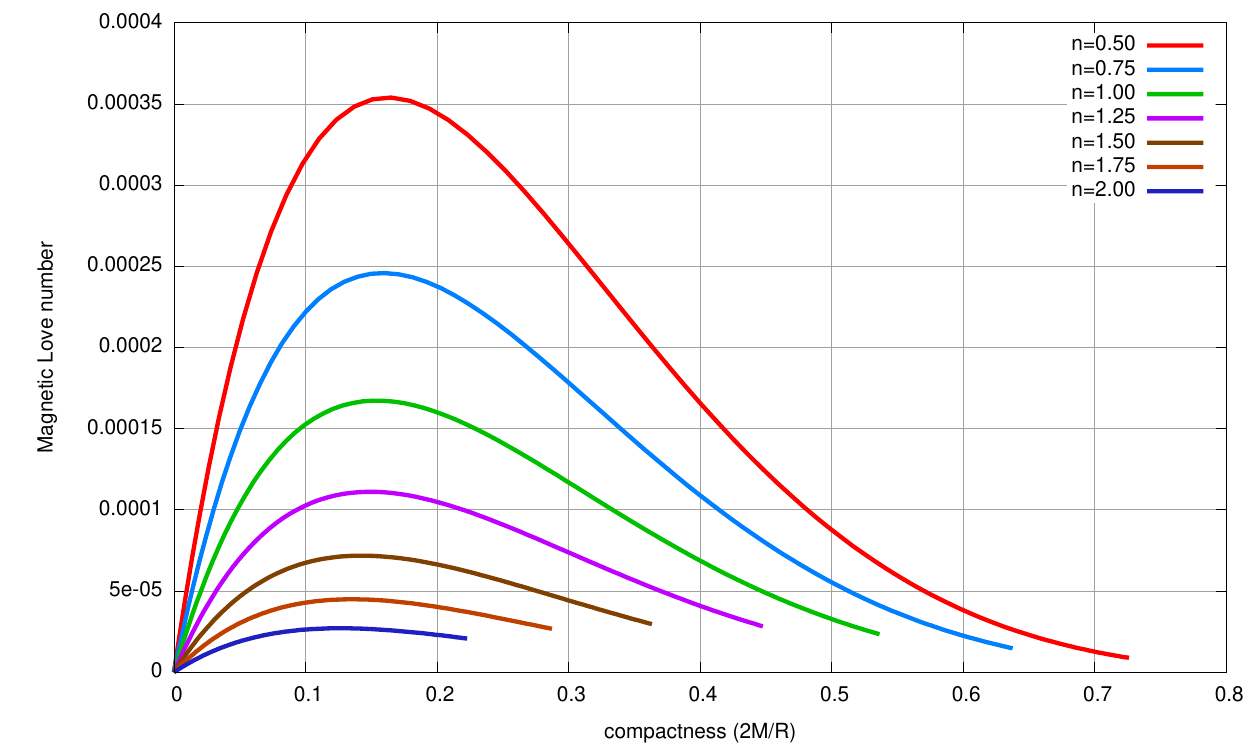}
\caption{Magnetic-type Love numbers for $l=5$.} 
\end{figure}

\subsection*{Damour and Nagar} 

After this work was completed we witnessed the appearance of an
article by Damour and Nagar \cite{damour-nagar:09} in which almost
identical work is presented. Their paper, like ours, is concerned with
the tidal deformation of compact bodies in full general relativity,
and presents precise definitions for electric-type and magnetic-type
Love numbers. And their paper, like ours, presents computations of
Love numbers for selected matter models. Their coverage of the
parameter space is wider: Damour and Nagar examine two types of
polytropic equations of state, and two tabulated equations of state for
realistic nuclear matter. In addition, Damour and Nagar define and
compute ``shape Love numbers,'' something that we did not pursue in 
this work.   

There are superficial differences between our treatments. One concerns
the choice of coordinates: Damour and Nagar work in
Schwarzschild coordinates and adopt the Regge-Wheeler gauge for the
metric perturbation; we work in Eddington-Finkelstein coordinates and
the light-cone gauge. Another concerns notation: we adopt different
normalization conditions for the Love numbers and the tidal
moments. These differences are not important.  

A more significant difference concerns the conclusion that the tidal
Love numbers of a black hole must be zero. In this paper we boldly
proclaim this conclusion, which we firmly believe to be a correct
interpretation of our results. Damour and Nagar, however, shy away
from the conclusion, although they agree with us on the basic
results. We do not understand the reasons behind this
reluctance. Damour and Nagar comment on the need to understand
``diverging diagrams that enter the computation of interacting black
holes at the 5-loop (or 5PN) level'' before reaching a conclusion. But
since the results presented here do not rely at all on a
post-Newtonian expansion of the field equations, the fate of 5PN terms
in a post-Newtonian representation of interacting black holes seems to
us to be irrelevant. We point out, also, that the Damour-Nagar work
does not provide a very clean foundation for the tidal deformation of
black holes, because their coordinate system is ill-behaved on the
event horizon. Our light-cone coordinates were selected precisely
because they permit a unified treatment of material bodies and black
holes.  

Aside from this issue of interpretation, and as far as we can judge,
the results presented here are in complete agreement with the
Damour-Nagar results. The Damour-Nagar work was carried out in
complete independence from us, and our work was carried out in
complete independence from them. The near-simultaneous completion of
our works provides evidence that the problem is interesting and
timely, and the agreement is a reassuring confirmation that each team
performed their calculations without error.      

\subsection*{Fang and Lovelace} 

The deformation of a black hole produced by an applied tidal field was 
previously examined by Fang and Lovelace \cite{fang-lovelace:05}, who 
concluded that $Q_{ab} = 0$ when the perturbation is expressed in
Regge-Wheeler gauge. Fang and Lovelace therefore anticipated our
result that the quadrupole, electric-type Love number of a black hole
is zero. These authors, however, qualified their conclusion by raising
doubts about the gauge invariance of the result, and claiming that the
induced quadrupole moment of a tidally deformed black hole is
inherently ambiguous. We do not share these reservations.  

We first discuss the issue of gauge invariance. The argument advanced
by Fang and Lovelace in favor of a gauge dependence of the tidal Love
number goes as follows. In Newtonian theory, the coordinate
transformation $r = \bar{r} [1+2\chi (R/\bar{r})^5]^{1/2}$, 
where $\chi$ is an arbitrary constant, turns a pure tidal potential
$\E_{ab} x^a x^b$ into $[1 + 2\chi (R/\bar{r})^5] 
\E_{ab} \bar{x}^a \bar{x}^b$, which appears to describe a sum of tidal
and body potentials; the transformation shifts the Love number by
$\chi$.  Fang and Lovelace correctly dismiss this coordinate
dependence as irrelevant in Newtonian theory, because $r$ has a 
well-defined meaning, but they point out that in a relativistic
context, the coordinate transformation could be viewed as a change of  
gauge. The implication,  then, is that the relativistic Love number
can be altered by a gauge transformation. Notice that the argument
applies to all types of compact bodies: material bodies and black
holes.   

We do not accept the validity of this argument. The coordinate
transformation considered by Fang and Lovelace is not of a type that
can be associated with a gauge transformation of the perturbation 
theory. A gauge transformation necessarily involves coordinate
displacements that are of the same order of magnitude as the
perturbation field. But the transformation from $r$ to $\bar{r}$ does
not involve the perturbation at all, and represents a large change of
the background coordinates. The new coordinate $\bar{r}$ does not
share the geometrical properties of the original $r$, and one would
easily be able to distinguish the two coordinate systems. The
argument, therefore, does not make a case for the gauge dependence of
the Love numbers. And in fact, the  {\it gauge invariance} of 
$k_{\rm el}$ and $k_{\rm mag}$ for all types of compact bodies
(material bodies and black holes) is established in Sec.~III.   

We next discuss the issue of ambiguity. Unlike Fang and Lovelace, we
believe that the relativistic Love numbers of compact bodies, as
defined in this paper, are well defined and completely devoid of
ambiguity. The reason is that the metric of
Eqs.~(\ref{deformed_metric}), which is presented in coordinates that
have clear geometrical properties, defines a perfectly well-defined
spacetime geometry. Given this spacetime, one could in principle
monitor the motion of test masses and  light rays and thereby measure
its detailed features, including the mass $M$, the tidal moments
$\E_L$ and $\B_L$, and the Love numbers. These measurements would
contain no ambiguities.    

The ambiguity identified by Fang and Lovelace concerns the coupling of 
$Q_{ab}$, the induced quadrupole moment, to $\E_{abc}$, the
octupole moment of the applied tidal field. According to Newtonian
ideas, this coupling should lead to a force  
$F^a = -\frac{1}{2} \E^a_{\ bc} Q^{bc}$ acting on the compact
body. (Once more the argument applies to all types of compact bodies.)  
Fang and Lovelace associate $F^a$ with $\dot{P}^a(r)$, the rate of
change of three-momentum contained within a world tube of
radius $r$ that surrounds the compact body; this is calculated by 
integrating the flux of Landau-Lifshitz energy-momentum pseudotensor 
across the world tube. They observe that the result is indeed
proportional to $\E^a_{\ bc} Q^{bc}$, but that the coefficient in
front depends on $r$. They interpret this as a statement that the
force is ambiguous, assign the ambiguity to $Q_{ab}$, and conclude
that the induced quadrupole moment of a tidally-deformed compact body
is inherently ambiguous. 

We believe that the ambiguity in $\dot{P}^a(r)$ is genuine --- the
result does depend on the world tube's radius. It is hasty, however,
to conclude from this that $F^a$ itself is ambiguous, because force   
calculations that rely on techniques of matched asymptotic expansions  
\cite{thorne-hartle:85, zhang:85} must involve a limiting procedure
in which both $M$ and $r$ are taken to approach zero. Although
ambiguities remain in this procedure, they are much smaller than those
claimed by Fang and Lovelace. At the accuracy level of our
calculations, the induced quadrupole moment of a tidally-deformed
compact body is not ambiguous.   

\subsection*{Suen}  

An earlier determination of the induced quadrupole moment of a tidally
deformed black hole was made by Suen \cite{suen:86b}, who examined the
specific case of a black hole perturbed by an axisymmetric ring of
matter. Suen found that the black-hole quadrupole moment 
is $Q_{ab} = +\frac{4}{21} M^5 \E_{ab}$, so that it gives rise to a
negative Love number, $k_{\rm el} = -\frac{1}{122}$. This result
contradicts our own results.  

Suen's result is wrong.  The starting points of Suen's analysis is the
perturbed metric presented in Eq.~(2.6) of his paper. It is easy to
show that while the metric does indeed satisfy the Einstein field
equations (up to terms that are quadratic in the small parameter $A$),
it fails to be regular at the event horizon. The metric does not,
therefore, represent a perturbed black hole, and the nonzero
result for $k_{\rm el}$ is a consequence of this fact. The regularity
of the metric perturbation $p_{\alpha\beta}$ at $r=2M$ can be judged
by examining its components in the light-cone coordinates
$(v,r,\theta,\phi)$, which are regular on the event horizon. A simple
calculation reveals that in Suen's notation, $p_{rr} = -2(2U-V)/f$,
where $f = 1-2M/r$. This is singular at $r=2M$ unless $2U-V$ vanishes
there, but Eqs.~(2.7) of Suen's paper show instead that $2U-V \to
AM^2$ in the limit. The perturbation is singular. 

\subsection*{Organization of the paper} 

In the remaining sections of this paper we present the details of our
analysis, and describe how the results reviewed previously were
obtained. We begin in Sec.~II with a discussion of tidal moments and 
tidal potentials, and motivate the definitions presented in
Eqs.~(\ref{Epot}) and (\ref{Bpot}). In Sec.~III we solve the external
problem, and show that the metric of Eqs.~(\ref{deformed_metric}) is a
solution to the vacuum field equations linearized about the
Schwarzschild metric. In Sec.~IV we formulate the internal problem for
general stellar models, and we specialize this to polytropes in
Sec.~V. In Sec.~VI we review the numerical techniques that were
employed to generate the figures and the tables displayed in the
Appendix.  

\section{Tidal moments and potentials} 

A spherical stellar model is perturbed by an external tidal field
characterized by the electric-type tidal moments $\E_L(v)$ and the
magnetic-type tidal moments $\B_L(v)$. These are symmetric-tracefree
(STF) tensors, and $L$ is a multi-index that contains a number $l$ of
individual indices. The tidal moments depend on $v$ (and not on the
spatial coordinates), but this time dependence is taken to be so slow
that all $v$-derivatives will be ignored in the Einstein field
equations.   

We begin our discussion of tidal potentials by adopting 
quasi-Cartesian coordinates $x^a$ related in the usual way to our
spherical coordinates $(r,\theta^A)$. We write the transformation as
$x^a = r\Omega^a(\theta^A)$, with $\Omega^a = [\sin\theta\cos\phi,
\sin\theta\sin\phi, \cos\theta]$ denoting the unit radial vector. We
introduce 
\begin{equation} 
\gamma_{ab} := \delta_{ab} - \Omega_a \Omega_b 
\label{projector} 
\end{equation} 
as the projector to the transverse space orthogonal to $\Omega^a$, and 
we let $\Omega^a_A := \partial \Omega^a/\partial \theta^A$. We note
the helpful identities  
\begin{subequations} \allowdisplaybreaks
\label{identities1}
\begin{align}  
\Omega_a \Omega^a_A &= 0, \\
\Omega_{AB} &= \gamma_{ab} \Omega^a_A \Omega^b_B =
\delta_{ab} \Omega^a_A \Omega^b_B, \\
\Omega^{AB} \Omega^a_A \Omega^b_B &= \gamma^{ab}. 
\end{align}
\end{subequations} 
Here $\Omega_{AB} = \mbox{diag}[1,\sin^2\theta]$ is the metric on the  
unit two-sphere, and $\Omega^{AB}$ is its inverse. We introduce $D_A$
as the covariant-derivative operator compatible with $\Omega_{AB}$,
and $\epsilon_{AB}$ as the Levi-Civita tensor on the unit two-sphere
(with nonvanishing components $\epsilon_{\theta\phi} =
-\epsilon_{\phi\theta} = \sin\theta$). In addition to
Eqs.~(\ref{identities1}) we also have  
\begin{subequations} \allowdisplaybreaks 
\label{identities2}
\begin{align} 
\epsilon_{AB} &= \epsilon_{abc} \Omega^a_A \Omega^b_B \Omega^c, \\ 
\epsilon_A^{\ B} \Omega^b_B &= -\Omega^a_A \epsilon_{ap}^{\ \ \ b}
\Omega^p, \\
D_A D_B \Omega^a &= D_B D_A \Omega^a = -\Omega^a \Omega_{AB}. 
\end{align}
\end{subequations} 
Here and below, upper-case latin indices are raised and lowered with
$\Omega^{AB}$ and $\Omega_{AB}$, respectively. Finally, we note that
$D_C \Omega_{AB} = D_C \epsilon_{AB} = 0$. 

For an electric-type tidal moment $\E_L$ of degree $l \geq 2$, the
Cartesian version of the tidal potentials are defined by  
\begin{subequations} \allowdisplaybreaks
\label{electric_potentialsC} 
\begin{align} 
\E^{(l)} &:= \E_L \Omega^L, \\ 
\E^{(l)}_a &:= \gamma_a^{\ c} \E_{c\,L-1} \Omega^{L-1}, \\ 
\E^{(l)}_{ab} &:= 2\gamma_a^{\ c} \gamma_b^{\ d} \E_{cd\, L-2} \Omega^{L-2} 
  + \gamma_{ab} \E^{(l)}. 
\end{align}
\end{subequations} 
Here $\E^{(l)}$ is a scalar potential, $\E^{(l)}_a$ is a transverse
vector potential, and $\E^{(l)}_{ab}$ is a transverse-tracefree tensor
potential. The angular version of the tidal potentials are 
\begin{subequations} \allowdisplaybreaks
\label{electric_potentialsS} 
\begin{align} 
\E^{(l)} &= \E_L \Omega^L, \\ 
\E^{(l)}_A &:= \E^{(l)}_a \Omega^a_A 
  = \Omega^a_A \E_{a\,L-1} \Omega^{L-1}, \\ 
\E^{(l)}_{AB} &:= \E^{(l)}_{ab} \Omega^a_A \Omega^b_B 
  = 2 \Omega^a_A \Omega^b_B \E_{ab\, L-2} \Omega^{L-2} 
  + \Omega_{AB} \E^{(l)}.
\end{align}
\end{subequations} 

For a magnetic-type tidal moment $\B_L$ of degree $l \geq 2$, the
Cartesian version of the tidal potentials are defined by  
\begin{subequations} \allowdisplaybreaks
\label{magnetic_potentialsC} 
\begin{align}  
\B^{(l)}_a &:= \epsilon_{apq} \Omega^p \B^q_{\ L-1} \Omega^{L-1}, \\ 
\B^{(l)}_{ab} &:= \bigl(\epsilon_{apq} \Omega^p 
  \B^q_{\ d\, L-2} \gamma^d_{\ b} 
  + \epsilon_{bpq} \Omega^p \B^q_{\ c\, L-2} 
  \gamma^c_{\ a} \bigr) \Omega^{L-2}. 
\end{align}
\end{subequations} 
Here $\B^{(l)}_a$ is a transverse vector potential, and
$\B^{(l)}_{ab}$ is a transverse-tracefree tensor potential; there is
no scalar potential in the magnetic case. The angular version of the
tidal potentials are  
\begin{subequations} \allowdisplaybreaks
\label{magnetic_potentialsS}  
\begin{align} 
\B^{(l)}_A &:= \B^{(l)}_a \Omega^a_A 
  = \Omega^a_A \epsilon_{apq} \Omega^p \B^q_{\ L-1} \Omega^{L-1}, \\  
\B^{(l)}_{AB} &:= \B^{(l)}_{ab} \Omega^a_A \Omega^b_B 
  = \bigl( \Omega^a_A \epsilon_{apq} \Omega^p \B^q_{\ b\, L-2}
  \Omega^b_B 
\nonumber \\ & \qquad \mbox{} 
+ \Omega^b_B \epsilon_{bpq} \Omega^p \B^q_{\ a\, L-2} 
  \Omega^a_A \bigr) \Omega^{L-2}.
\end{align} 
\end{subequations} 

The tidal potentials can all be expressed in terms of (scalar, vector,
and tensor) spherical harmonics. Let $Y^{lm}$ be the standard (scalar)  
spherical-harmonic functions. The vector and tensor harmonics of
even parity are $Y^{lm}_A := D_A Y^{lm}$, $\Omega_{AB} Y^{lm}$, and  
$Y^{lm}_{AB} := [ D_A D_B + \frac{1}{2} l(l+1) \Omega_{AB}] Y^{lm}$;  
notice that $\Omega^{AB} Y^{lm}_{AB} = 0$ by virtue of the eigenvalue
equation satisfied by the spherical harmonics. The vector and
tensor harmonics of odd parity are $X^{lm}_A := -\epsilon_A^{\ B}
D_B Y^{lm}$ and $X^{lm}_{AB} := -\frac{1}{2} ( \epsilon_A^{\ C} D_B 
+ \epsilon_B^{\ C} D_A ) D_C Y^{lm}$; $X^{lm}_{AB}$ also is tracefree: 
$\Omega^{AB} X^{lm}_{AB} = 0$.   

We first express the electric-type tidal potentials in terms of the
even-parity spherical harmonics. We begin with $\E^{(l)}$, which we
decompose as   
\begin{equation} 
\E^{(l)}(v,\theta^A) = \sum_m \E^{(l)}_m(v) Y^{lm}(\theta^A), 
\label{scalar_E} 
\end{equation}
in terms of harmonic components $\E^{(l)}_m(v)$. There are $2l+1$ terms
in the sum, and the $2l+1$ independent components of $\E_L$ are in a
one-to-one correspondence with the $2l+1$ coefficients
$\E^{(l)}_m$. Returning to the original representation of
Eq.~(\ref{electric_potentialsC}), we find after differentiation that
$D_A \E^{(l)} = l \Omega^a_A \E_{a\, L-1} \Omega^{L-1}$, and we
conclude that  
\begin{equation} 
\E^{(l)}_A = \frac{1}{l} D_A \E^{(l)} 
= \frac{1}{l} \sum_m \E^{(l)}_m Y_A^{lm}. 
\label{vector_E} 
\end{equation} 
An additional differentiation using the last of
Eqs.~(\ref{identities2}) reveals that $D_A D_B \E^{(l)} = -l
\Omega_{AB} \E^{(l)} + l(l-1) \Omega^a_A \Omega^b_B 
\E_{ab\, L-2} \Omega^{L-2}$. From this we conclude that  
\begin{eqnarray} 
\E^{(l)}_{AB} &=& \frac{2}{l(l-1)} \Bigl[ D_A D_B 
+ \frac{1}{2} l(l+1) \Omega_{AB} \Bigr] \E^{(l)}
\nonumber \\  
&=& \frac{2}{l(l-1)} \sum_m \E^{(l)}_m Y_{AB}^{lm}. 
\label{tensor_E}
\end{eqnarray} 

We next express the magnetic-type potentials in terms of the
odd-parity spherical harmonics. We begin with $\B^{(l)} := 
\B_L \Omega^L$ and its decomposition $\B^{(l)} = \sum_m 
\B^{(l)}_m Y^{lm}(\theta^A)$. Differentiating the first expression,
multiplying this by the Levi-Civita tensor, and involving the second
of Eqs.~(\ref{identities2}) returns $\epsilon_A^{\ B} D_B \B^{(l)} =
-l \Omega^a_A \epsilon_{apq} \Omega^p \B^q_{\ L-1} 
\Omega^{L-1}$. From this we conclude that 
\begin{equation} 
\B^{(l)}_A = \frac{1}{l} \bigl( -\epsilon_A^{\ B} D_B \bigr) \B^{(l)}  
= \frac{1}{l} \sum_m \B^{(l)}_m X_A^{lm}. 
\label{vector_B}
\end{equation} 
A second differentiation yields 
$-\epsilon_A^{\ C} D_B D_C \B^{(l)} = l \epsilon_{AB} \B^{(l)} 
+ l(l-1) \Omega^a_A \epsilon_{apq} \Omega^p \B^q_{\ b\, L-2}
\Omega^b_B \Omega^{L-2}$, and after symmetrization we obtain 
\begin{eqnarray} 
\B^{(l)}_{AB} &=& -\frac{1}{l(l-1)} \bigl( \epsilon_A^{\ C} D_B 
+ \epsilon_B^{\ C} D_A \bigr) D_C \B^{(l)}  
\nonumber \\
&=& \frac{2}{l(l-1)} \sum_m \B^{(l)}_m X_{AB}^{lm}. 
\label{tensor_B}
\end{eqnarray} 

\section{External problem} 

\subsection{Even-parity sector} 

In this subsection we determine the tidal deformation of the metric  
outside the matter distribution, in the even-parity sector. The
unperturbed external solution is the Schwarzschild metric 
\begin{equation} 
ds_0^2 = -f\, dv^2 + 2 dvdr + r^2\, d\Omega^2,
\label{Schwarzschild} 
\end{equation} 
with $f := 1 - 2M/r$ and $M$ denoting the body's mass; the metric is 
valid for $r > R$, where $R$ is the body's radius. We employ the
perturbation formalism of Martel and Poisson \cite{martel-poisson:05},
and implement the light-cone gauge of Preston and Poisson
\cite{preston-poisson:06b}.   

In the light-cone gauge the even-parity metric perturbation is given
by 
\begin{subequations} \allowdisplaybreaks 
\label{even_perturbation1} 
\begin{align}  
p_{vv} &= \sum_m h^{lm}_{vv}(r) Y^{lm}(\theta^A), \\ 
p_{vA} &= \sum_m j^{lm}_v(r) Y^{lm}_A(\theta^A), \\ 
p_{AB} &= r^2 \sum_m K^{lm}(r) \Omega_{AB} Y^{lm}(\theta^A) 
\nonumber \\ & \qquad \mbox{}   
+ r^2 \sum_m G^{lm}(r) Y^{lm}_{AB}(\theta^A). 
\end{align}
\end{subequations} 
We consider each $l$-mode separately, and we henceforth omit
the label $lm$ on the perturbation variables $h_{vv}$, $j_v$, $K$, and
$G$, which depend on $r$ only. As discussed by Preston and Poisson,
$K^{lm}$ can always be set equal to zero when the perturbation
satisfies the vacuum field equations; this represents a refinement of
the light-cone gauge, and we shall make this choice here. 

To simplify the task of solving the field equations we set 
\begin{subequations} \allowdisplaybreaks 
\label{even_perturbation2} 
\begin{align} 
h_{vv} &= -\frac{2}{(l-1)l} r^l e_1(r) \E^{(l)}_m, \\ 
j_{v} &= -\frac{2}{(l-1)l(l+1)} r^{l+1} e_4(r) \E^{(l)}_m, \\ 
G &= -\frac{4}{(l-1)l^2(l+1)} r^l e_7(r) \E^{(l)}_m,
\end{align}
\end{subequations} 
where the functions $e_1(r)$, $e_4(r)$, and $e_7(r)$ are to be
determined. Substitution of Eqs.~(\ref{even_perturbation2}) into
Eq.~(\ref{even_perturbation1}) produces 
\begin{subequations} \allowdisplaybreaks 
\label{even_perturbation3}
\begin{align}  
p_{vv} &= -\frac{2}{(l-1)l} r^l e_1(r) \E^{(l)}, \\ 
p_{vA} &= -\frac{2}{(l-1)(l+1)} r^{l+1} e_4(r) \E^{(l)}_A, \\
p_{AB} &= -\frac{2}{l(l+1)} r^{l+2} e_7(r) \E^{(l)}_{AB},
\end{align}
\end{subequations} 
where $\E^{(l)}$, $\E^{(l)}_A$, and $\E^{(l)}_{AB}$ are the tidal
potentials introduced in Eq.~(\ref{electric_potentialsS}). 

The motivation behind the introduction of the functions $e_1$, $e_4$,
and $e_7$ goes as follows. We first observe that when we set $e_1 =
e_4 = e_7 = 1$, the perturbation defined by
Eqs.~(\ref{even_perturbation2}) or Eqs.~(\ref{even_perturbation3})
satisfies the equations of linearized theory for a perturbation of
Minkowski spacetime. This exercise reveals that $h_{vv}$ must be
proportional to $r^l$, $j_v$ to $r^{l+1}$, and $G$ to $r^l$; the
relative numerical coefficients between these fields are also
determined by solving the perturbation equations in flat
spacetime. The remaining absolute numerical coefficient that relates
the perturbation to the tidal moment $\E_L$ is determined by the 
definition of the tidal moment in terms of the Weyl tensor of the
perturbed spacetime; this coefficient --- the factor $-2/[(l-1)l]$ in
$h_{vv}$ --- can be read off Eq.~(3.26a) of Ref.~\cite{zhang:86}.   

Inserting the functions $e_1$, $e_4$, and $e_7$ in
Eqs.~(\ref{even_perturbation2}) allows the perturbation to be a
solution to the Einstein field equations linearized about the
Schwarzschild metric instead of the Minkowski metric. We impose the
boundary conditions  
\begin{equation} 
e_1(r\to\infty) = e_4(r\to\infty) = e_7(r\to\infty) = 1. 
\label{even_boundary} 
\end{equation} 
The field equations do not determine these functions uniquely. The
light-cone gauge comes with a class of residual gauge transformations  
that preserve the light-cone nature of the coordinate system (see
Preston and Poisson \cite{preston-poisson:06b}). In the even-parity
sector, and for static perturbations, the residual gauge freedom that
keeps $K=0$ is a one-parameter family described by  
\begin{subequations} \allowdisplaybreaks 
\label{even_gauge} 
\begin{align} 
e_1 &\to e_1 - l a_1 (2M/r)^{l+2}, \\ 
e_4 &\to e_4 + a_1 \bigl[ (l-1)(l+2) + 4M/r \bigr] (2M/r)^{l+1}, \\  
e_7 &\to e_7 + 2l a_1 (2M/r)^{l+1},
\end{align}
\end{subequations}
in which $a_1$ is the (dimensionless) parameter. The residual gauge 
freedom does not interfere with the boundary conditions of
Eq.~(\ref{even_boundary}). 

When $K$ is allowed to change, the residual gauge freedom becomes a
three-parameter family. In this case we have 
\begin{subequations} \allowdisplaybreaks 
\label{even_gauge_withK} 
\begin{align} 
e_1 &\to e_1 - l a_1 (2M/r)^{l+2} + a_3 (2M/r)^{l+2}, \\ 
e_4 &\to e_4 + a_1 \bigl[ (l-1)(l+2) + 4M/r \bigr] (2M/r)^{l+1} 
\nonumber \\ & \qquad \mbox{}
- (l+1) a_3 (2M/r)^{l+1}, \\ 
e_7 &\to e_7 + 2l a_1 (2M/r)^{l+1} + 2a_2 (2M/r)^l, 
\end{align}
\end{subequations}
and $K$ becomes 
\begin{equation} 
K = \frac{ 4(2M)^l }{(l-1)l} \bigl[ a_2 + a_3 (2M/r) \bigr]
\E^{(l)}_m. 
\label{K}
\end{equation} 
Here $a_2$ and $a_3$ are two additional gauge parameters.  

The differential equations satisfied by $e_1$, $e_4$, and $e_7$ can be
extracted from the perturbation equations. These equations are
coupled, and some effort must be devoted to their decoupling before an
attempt is made to find solutions. We shall not describe these routine
steps here. We state simply that the solutions are the ones that were
displayed in Eqs.~(\ref{radial}) and (\ref{functions}). These are
given in a minimal implementation of the light-cone gauge, in which
all constants of integrations are set equal to zero. The most general
form of the solution is obtained from this by effecting the shifts
described by Eqs.~(\ref{even_gauge_withK}) and (\ref{K}). The
functions $A_n$ and $B_n$ are displayed for selected values of $l$ in
Table I.    

\squeezetable
\begin{table*}
\caption{Functions $A_n$ and $B_n$ for selected values of $l$,
  expressed in terms of $z := 2M/r$. The numbers $\mu_l$ and 
  $\lambda_l$ are given by $\lambda_l = 
  (2l)!(2l+1)!/[(l-2)!(l-1)!(l+1)!(l+2)!]$ and 
  $\mu_l = (l+1)\lambda_l/l$.}  
\begin{ruledtabular} 
\begin{tabular}{ll} 
$l = 2$ & $\mu_2 = 30$, \qquad $\lambda_2 = 20$ \\ 
$A_1 = (1-z)^2$ & 
$z^5 B_1 = -\mu_2 A_1 \ln(1-z) - \frac{5}{2} z(2-z)(6-6z-z^2)$ \\ 
$A_4 = 1-z$ & 
$z^5 B_4 = \lambda_2 A_4 \ln(1-z) + \frac{5}{3} z(12-6z-2z^2-z^3)$ \\ 
$A_7 = 1-\frac{1}{2} z^2$ & 
$z^5 B_7 = -\mu_2 A_7 \ln(1-z) - 5z(6+3z-z^2)$ \\ 
\hline
$l = 3$ & $\mu_3 = 840$, \qquad $\lambda_3 = 630$ \\ 
$A_1 = \frac{1}{2}(1-z)^2(2-z)$ & 
$z^7 B_1 = -\mu_3 A_1 \ln(1-z) - 7z(120-240z+130z^2-10z^3-z^4)$ \\
$A_4 = \frac{1}{3}(1-z)(3-2z)$ & 
$z^7 B_4 = \lambda_3 A_4 \ln(1-z) 
  + \frac{7}{2} z(180-210z+30z^2+5z^3+z^4)$ \\ 
$A_7 = 1-z+\frac{1}{10}z^3$ & 
$z^7 B_7 = -\mu_3 A_7 \ln(1-z) - 14z(60-30z-10z^2+z^3)$ \\ 
\hline 
$l = 4$ & $\mu_4 = 17\ 640$, \qquad $\lambda_4 = 14\ 112$ \\ 
$A_1 = \frac{1}{14}(1-z)^2(14-14z+3z^2)$ & 
$z^9 B_1 = -\mu_4 A_1 \ln(1-z) - 21z(2-z)(420-840z+440z^2-20z^3-z^4)$ \\ 
$A_4 = \frac{1}{28}(1-z)(28-35z+10z^2)$ & 
$z^9 B_4 = \lambda_4 A_4 \ln(1-z) + \frac{42}{5}
  z(1680-2940z+1370z^2-90z^3-9z^4-z^5)$ \\ 
$A_7 = 1-\frac{5}{3}z+\frac{5}{7}z^2-\frac{1}{42}z^4$ &
$z^9 B_7 = -\mu_4 A_7 \ln(1-z) - 14z(1260-1470z+270z^2+65z^3-3z^4)$ \\     
\hline
$l = 5$ & $\mu_5 = 332\ 640$, \qquad $\lambda_5 = 277\ 200$ \\ 
$A_1 = \frac{1}{12}(1-z)^2(2-z)(6-6z+z^2)$ & 
$z^{11} B_1 = -\mu_5 A_1 \ln(1-z) 
  - 66z(5040-15120z+16380z^2-7560z^3+1288z^4-28z^5-z^6)$ \\
$A_4 = \frac{1}{30}(1-z)(30-54z+30z^2-5z^3)$ & 
$z^{11} B_4 = \lambda_5 A_4 \ln(1-z) 
+ 22z(12600-28980z+21840z^2-5670z^3+210z^4+14z^5+z^6)$ \\ 
$A_7 = 1-\frac{9}{4}z+\frac{5}{3}z^2-\frac{5}{12}z^3+\frac{1}{168}z^5$
& $z^{11} B_7 = -\mu_5 A_7 \ln(1-z) 
  - 66z(5040-8820z+4410z^2-420z^3-77z^4+2z^5)$ 
\end{tabular}
\end{ruledtabular}
\end{table*} 

The metric perturbation can be represented in terms of gauge-invariant 
variables. We employ the set defined by Eqs.~(4.10)--(4.12) of Martel
and Poisson \cite{martel-poisson:05}. According to these equations,
and as can be directly verified from Eq.~(\ref{even_gauge_withK}), the
variables 
\begin{subequations} \allowdisplaybreaks 
\label{even_gaugeinv1} 
\begin{align} 
\tilde{h}_{vv} &:= h_{vv} + \frac{2M}{r^2} j_v - Mf G', \\ 
\tilde{h}_{vr} &:= MG' - j'_v, \\ 
\tilde{h}_{rr} &:= 2r G' + r^2 G'', \\ 
\tilde{K} &:= -\frac{2}{r} j_v + \frac{1}{2} l(l+1) G + rf G'
\end{align}
\end{subequations} 
are gauge-invariant; a prime indicates differentiation with respect to
$r$. We express them as 
\begin{subequations} \allowdisplaybreaks 
\label{even_gaugeinv2} 
\begin{align} 
\tilde{h}_{vv} &:= -\frac{2}{(l-1)l} r^l e_{vv}(r) \E^{(l)}_m, \\  
\tilde{h}_{vr} &:= \frac{2}{(l-1)l} r^l e_{vr}(r) \E^{(l)}_m, \\ 
\tilde{h}_{rr} &:= -\frac{4}{(l-1)l} r^l e_{rr}(r) \E^{(l)}_m, \\ 
\tilde{K} &:= -\frac{2}{(l-1)l} r^l e_{K}(r) \E^{(l)}_m,
\end{align}
\end{subequations} 
in terms of new radial functions $e_{vv}$, $e_{vr}$, $e_{rr}$, and
$e_K$. Calculation reveals that these are given in terms of the old
ones by 
\begin{subequations} \allowdisplaybreaks 
\label{even_gaugeinv3} 
\begin{align} 
e_{vv} &= e_1 + \frac{1}{l+1} \frac{2M}{r} e_4 
- \frac{1}{l+1} \frac{2M}{r} f e_7 
\nonumber \\ & \qquad \mbox{} 
- \frac{1}{l(l+1)} 2Mf e'_7, \\ 
e_{vr} &= e_4 + \frac{1}{l+1} r e'_4 - \frac{1}{l+1} \frac{2M}{r} e_7 
\nonumber \\ & \qquad \mbox{} 
- \frac{1}{l(l+1)} 2M e'_7, \\ 
e_{rr} &= e_7 + \frac{2}{l} r e'_7 + \frac{1}{l(l+1)} r^2 e''_7, \\ 
e_{K} &= -\frac{2}{l+1} e_4 + \frac{1}{l+1} (l+3-4M/r) e_7 
\nonumber \\ & \qquad \mbox{} 
+ \frac{2}{l(l+1)} r f e'_7. 
\end{align}
\end{subequations} 
It is easy to see that these functions, like the old ones, all go to
one as $r$ goes to infinity. 
 
Substitution of our expressions for $e_1$, $e_4$, and $e_7$ into 
Eqs.~(\ref{even_gaugeinv3}) and repeated use of the properties of
hypergeometric functions reveal that  
\begin{equation} 
e_{vv} = f e_{vr} = f^2 e_{rr} = A_1 + 2k_{\rm el} (R/r)^{2l+1} B_1
\label{even_gaugeinv4} 
\end{equation} 
and 
\begin{equation} 
e_K = A_7 + 2 k_{\rm el} (R/r)^{2l+1} B_7.  
\label{even_gaugeinv5} 
\end{equation} 
Notice that $e_{vv}$, $fe_{vr}$, and $f^2 e_{rr}$ are all equal to the
minimal implementation of $e_1$, and $e_K$ is equal to the minimal
implementation of $e_7$. All this shows that the relativistic Love
numbers $k_{\rm el}$ possess gauge-invariant significance.  

\subsection{Odd-parity sector} 

In the light-cone gauge the odd-parity metric perturbation is given 
by 
\begin{subequations} \allowdisplaybreaks 
\label{odd_perturbation1} 
\begin{align} 
p_{vA} &= \sum_m h^{lm}_v(r) X^{lm}_A(\theta^A), \\
p_{AB} &= \sum_m h^{lm}_2(r) X^{lm}_{AB}(\theta^A). 
\end{align} 
\end{subequations} 
We consider each $l$-mode separately, and we henceforth omit
the label $lm$ on the perturbation variables $h_v$ and $h_2$, which
depend on $r$ only. To simplify the task of solving the field
equations we set  
\begin{subequations} \allowdisplaybreaks 
\label{odd_perturbation2} 
\begin{align} 
h_v &= \frac{2}{3(l-1)l} r^{l+1} b_4(r) \B^{(l)}_m, \\ 
h_2 &= \frac{4}{3(l-1)l^2} r^{l+2} b_7(r) \B^{(l)}_m,
\end{align}
\end{subequations} 
where the functions $b_4(r)$ and $b_7(r)$ are to be
determined. Substitution of Eqs.~(\ref{odd_perturbation2}) into
Eq.~(\ref{odd_perturbation1}) produces 
\begin{subequations} \allowdisplaybreaks 
\label{odd_perturbation3}
\begin{align}  
p_{vA} &= \frac{2}{3(l-1)} r^{l+1} b_4(r) \B^{(l)}_A, \\
p_{AB} &= \frac{2}{3l} r^{l+2} b_7(r) \B^{(l)}_{AB},
\end{align}
\end{subequations} 
where $\B^{(l)}_A$ and $\B^{(l)}_{AB}$ are the tidal
potentials first introduced in Eq.~(\ref{magnetic_potentialsS}). 

The motivation behind the introduction of the functions $b_4$ and 
$b_7$ is identical to what was done in the even-parity sector. 
When we set $b_4 = b_7 = 1$, the perturbation defined by
Eqs.~(\ref{odd_perturbation2}) or Eqs.~(\ref{odd_perturbation3})
satisfies the equations of linearized theory for a perturbation of
Minkowski spacetime. This exercise reveals the relative numerical
coefficients between $h_v$ and $h_2$. The remaining absolute numerical 
coefficient that relates the perturbation to the tidal moment $\B_L$
is determined by the definition of the tidal moment in terms of the
Weyl tensor of the perturbed spacetime; this coefficient --- the
factor $2/[3(l-1)l]$ in $h_v$ --- can be read off Eq.~(3.26b) of
Ref.~\cite{zhang:86}.    

Inserting the functions $b_4$ and $b_7$ in
Eqs.~(\ref{odd_perturbation2}) allows the perturbation to be a
solution to the Einstein field equations linearized about the
Schwarzschild metric instead of the Minkowski metric. We impose the
boundary conditions  
\begin{equation}  
b_4(r \to \infty) = b_7(r \to \infty) = 1. 
\label{odd_boundary} 
\end{equation} 
The field equations do not determine these functions uniquely. As in
the even-parity case we have a residual gauge freedom that preserves
the nature of the light-cone coordinates. It is described by 
\begin{subequations} \allowdisplaybreaks 
\label{odd_gauge} 
\begin{align} 
b_4 &\to b_4 \\ 
b_7 &\to b_7 + \alpha \Bigl( \frac{2M}{r} \Bigr)^l,
\end{align}
\end{subequations}
in which $\alpha$ is a (dimensionless) parameter. The residual gauge  
freedom does not interfere with the boundary conditions of
Eq.~(\ref{odd_boundary}). 

The differential equations satisfied by $b_4$ and $b_7$ can be
extracted from the perturbation equations. The solutions are displayed
in Eqs.~(\ref{radial}) and (\ref{functions}). They are given in a
minimal implementation of the light-cone gauge, in which all constants
of integrations are set equal to zero. The most general form of the
solution is obtained from this by effecting the shifts described by
Eqs.~(\ref{odd_gauge}). 

The metric perturbation can be represented in terms of
gauge-invariant variables. We employ the set defined by Eq.~(5.7) of
Martel and Poisson \cite{martel-poisson:05}. According to this, and as
can be directly verified from Eq.~(\ref{odd_gauge}), the variables
\begin{subequations} \allowdisplaybreaks 
\label{odd_gaugeinv1} 
\begin{align} 
\tilde{h}_v &:= h_v, \\ 
\tilde{h}_r &:= \frac{1}{r} h_2 - \frac{1}{2} h_2' 
\end{align}
\end{subequations} 
are gauge-invariant. We express them as 
\begin{subequations} \allowdisplaybreaks 
\label{odd_gaugeinv2} 
\begin{align} 
\tilde{h}_v &:= \frac{2}{3(l-1)l} r^{l+1} b_v(r) \B^{(l)}_m, \\  
\tilde{h}_r &:= -\frac{2}{3(l-1)l} r^{l+1} b_r(r) \B^{(l)}_m, \\  
\end{align}
\end{subequations} 
in terms of new radial functions $b_v$ and $b_r$. Calculation reveals
that these are given in terms of the old ones by 
\begin{subequations} \allowdisplaybreaks 
\label{odd_gaugeinv3} 
\begin{align} 
b_v &= b_4, \\ 
b_r &= b_7 + \frac{r}{l} b_7'. 
\end{align}
\end{subequations} 
It is easy to see that these functions, like the old ones, all go to
one as $r$ goes to infinity. 
 
Substitution of our expressions for $b_4$ and $b_7$ into
Eqs.~(\ref{odd_gaugeinv3}) and repeated use of the properties of the
hypergeometric functions reveal that 
\begin{equation} 
b_v = f b_r = A_4 - 2\frac{l+1}{l} k_{\rm mag} (R/r)^{2l+1} B_4.
\label{odd_gaugeinv4} 
\end{equation} 
Notice that $b_v$ and $f b_r$ are both equal to $b_4$, which is
gauge-invariant. This shows that the relativistic Love numbers 
$k_{\rm mag}$ possess gauge-invariant significance.  

\section{Internal problem} 

\subsection{Background metric for relativistic stellar models}  

We begin with an examination of the internal gravitational field of a 
body that is not yet perturbed by an external tidal field. The body
is spherically-symmetric, and the matter consists of a perfect fluid. 
In light-cone coordinates $(v,r,\theta^A)$ the metric is expressed as   
\begin{equation} 
ds_0^2 = -e^{2\psi} f\, dv^2 + 2e^\psi\, dvdr + r^2\, d\Omega^2,
\label{metric} 
\end{equation} 
with $f = 1 - 2m(r)/r$ and $\psi = \psi(r)$. The Einstein field
equations are  
\begin{equation} 
m' = 4\pi r^2 \rho, \qquad 
\psi' = \frac{4\pi r}{f} (\rho + p), 
\label{Einstein} 
\end{equation} 
and the equation of hydrostatic equilibrium is 
\begin{equation} 
p' = - \frac{m + 4\pi r^3 p}{r^2 f} (\rho + p). 
\label{TOV} 
\end{equation} 
Here $\rho$ is the fluid's proper energy density, and $p$ is the 
pressure. 

These equations can be integrated once an equation of state is
specified. The boundary conditions are $m(r=0) = 0$ and 
$\psi(r=0) = \psi_0$, where $\psi_0$ is chosen so that $\psi$ vanishes
at the stellar surface: $\psi(r=R) = 0$.   
 
\subsection{Light-cone gauge} 

The internal light-cone gauge is a modified version of the external
gauge constructed by Preston and Poisson
\cite{preston-poisson:06b}. We define it properly in this section.   

The metric of Eq.~(\ref{metric}) reveals the meaning of the
coordinates $(v,r,\theta^A)$ in the background spacetime. We note
first that $l_\alpha = -\partial_\alpha v$ is a null vector, so that
the surfaces $v = \mbox{constant}$ are null hypersurfaces; they
describe light cones that converge toward $r=0$. The vector
\begin{equation} 
l^\alpha = (0, -e^{-\psi}, 0, 0) 
\end{equation}
is tangent to the null generators of these light cones, and the
expression reveals that $\theta^A$ is constant along the
generators. In addition, the affine parameter $\lambda$ that runs
along the generators is related to $r$ by $d\lambda = -e^\psi\, dr$. 
In the interior portion of the spacetime, $r$ is no longer an affine
parameter on the null generators; but it still possesses the property
of being an areal radius, in the sense that the area of a surface of
constant $(v,r)$ is given by $4\pi r^2$.  

In the {\it internal light-cone gauge}, the metric of the perturbed
spacetime is presented in coordinates
$(v,r,\theta^A)$ that possess the same geometrical meaning as in the
background spacetime. In particular, $v$ continues to label null
hypersurfaces, $\theta^A$ continues to be constant along the null
generators, and $r$ continues to be related to the affine parameter
by $d\lambda = -e^\psi\, dr$. It is easy to show that these statements
imply the same conditions  
\begin{equation} 
p_{vr} = p_{rr} = p_{rA} = 0 
\label{int_lightconegauge} 
\end{equation} 
that were employed in the external problem. The nonvanishing
components of the metric perturbation are therefore $p_{vv}$,
$p_{vA}$, and $p_{AB}$. The radial coordinate, however, will lose its
meaning as an areal radius in the stellar interior.   

In the even-parity sector the perturbation is decomposed as in
Eq.~(\ref{even_perturbation1}), and the fields $h^{lm}_{vv}$,
$j^{lm}_{v}$, $K^{lm}$, $G^{lm}$ depend (in general) on the
coordinates $(v,r)$. An even-parity gauge transformation is generated
by the vector field $\Xi_\alpha$, with components  
\begin{subequations} \allowdisplaybreaks
\label{even_parity_gauge} 
\begin{align}  
\Xi_v &= \sum_{lm} \xi^{lm}_v(v,r) Y^{lm}(\theta^A), \\ 
\Xi_r &= \sum_{lm} \xi^{lm}_r(v,r) Y^{lm}(\theta^A), \\ 
\Xi_A &= \sum_{lm} \xi^{lm}(v,r) Y^{lm}_A(\theta^A). 
\end{align} 
\end{subequations} 
It can be shown that the condition $h_{vr} = 0$ determines $\xi_v$,
that $h_{rr} = 0$ determines $\xi_r$, and that $j_r = 0$ determines
$\xi$. The gauge, however, is not determined uniquely. There exists a
residual gauge freedom that preserves the geometrical meaning of the
coordinates. In the case of $v$-independent perturbations, the
residual gauge freedom is a three-parameter family described by 
\begin{subequations} \allowdisplaybreaks 
\label{even_residual} 
\begin{align} 
\xi_v &= -a_1 e^{2\psi} f + a_2, \\ 
\xi_r &= a_1 e^\psi, \\ 
\xi &= -a_1 r^2 \int^r r^{\prime -2} e^{\psi(r')}\, dr' + a_3 r^2. 
\end{align} 
\end{subequations} 
Here we suppressed the $lm$ labels on $\xi_v$, $\xi_r$, and $\xi$, as
well as the constants $a_1$, $a_2$, and $a_3$. 

In the odd-parity sector the perturbation is decomposed as in
Eq.~(\ref{odd_perturbation1}), and the fields $h^{lm}_v$, $h^{lm}_2$
depend (in general) on the coordinates $(v,r)$. An odd-parity gauge
transformation is generated by the vector field $\Xi_\alpha$, with
components 
\begin{equation} 
\Xi_v = \Xi_r = 0, \qquad 
\Xi_A = \sum_{lm} \xi^{lm}(v,r) X^{lm}_A(\theta^A). 
\label{odd_parity_gauge} 
\end{equation} 
It can be shown that the condition $h_{r} = 0$ determines $\xi$. In
this case also there exists a residual gauge freedom that preserves
the geometrical meaning of the coordinates. In the case of
$v$-independent perturbations, the residual gauge freedom is a
one-parameter family described by  
\begin{equation} 
\xi = \alpha r^2. 
\label{odd_residual} 
\end{equation} 
Here also we suppressed the $lm$ labels on $\xi$ and the constant  
$\alpha$. 

The decompositions of Eq.~(\ref{even_perturbation1}) and
(\ref{odd_perturbation1}) can be used to compute $\delta
G_{\alpha\beta}$, the perturbation of the Einstein tensor inside the
body. The even-parity sector decouples from the odd-parity sector, 
and the perturbation takes the form of 
\begin{subequations} \allowdisplaybreaks 
\label{pert_Einstein} 
\begin{align} 
\delta G_{vv} &= \sum_{lm} A^{lm}_{vv} Y^{lm}, \\ 
\delta G_{vr} &= \sum_{lm} A^{lm}_{vr} Y^{lm}, \\ 
\delta G_{rr} &= \sum_{lm} A^{lm}_{rr} Y^{lm}, \\ 
\delta G_{vA} &= \sum_{lm} \bigl( A^{lm}_v Y^{lm}_A 
  + B^{lm}_v X^{lm}_A \bigr), \\ 
\delta G_{rA} &= \sum_{lm} \bigl( A^{lm}_r Y^{lm}_A 
  + B^{lm}_r X^{lm}_A \bigr), \\ 
\delta G_{AB} &= \sum_{lm} \bigl( A^{lm}_\flat \Omega_{AB} Y^{lm} 
  + A^{lm}_\sharp Y_{AB} + B^{lm} X^{lm}_{AB} \bigr). 
\end{align}
\end{subequations} 
Here the even-parity fields $A_{vv}$, $A_{vr}$, $A_{rr}$, $A_v$,
$A_r$, $A_\flat$, $A_\sharp$ and the odd-parity fields $B_v$, $B_r$,
$B$ depend on $v$ and $r$ only. In the case of a stationary
perturbation, they depend on $r$ only.   

The expressions are too long to be displayed here. In practice they
are easily generated with GRTensorII \cite{grtensor} by specializing
the perturbation to an axisymmetric mode $m=0$ with a specific
multipole order $l$. With $Y^{lm} = Y(\theta)$ we have 
$Y_\theta = Y'$, $Y_\phi = 0$, 
$Y_{\theta\theta} = -\cos\theta Y'/\sin\theta - \frac{1}{2}
l(l+1) Y$, $Y_{\theta\phi} = 0$, and $Y_{\phi\phi} 
= \sin\theta\cos\theta Y' + \frac{1}{2} l(l+1) \sin^2\theta Y$ in the
even-parity case, and $X_\theta = 0$, $X_\phi = \sin\theta Y'$,
$X_{\theta\theta} = 0$, $X_{\theta\phi} = -\cos\theta Y' - \frac{1}{2}
l(l+1) \sin\theta Y$, and $X_{\phi\phi} = 0$ in the odd-parity
case. The definition of the metric implements the constraint $Y'' =
-\cos\theta Y'/\sin\theta - l(l+1) Y$ on the spherical-harmonic
functions, and this simplifies the final expression for the perturbed
Einstein tensor. 

\subsection{Energy-momentum tensor} 

We consider {\it stationary tides} raised by a tidal environment 
characterized by an electric-type tidal moment $\E_L$ and a
magnetic-type tidal moment $\B_L$; these are actually time-dependent, 
but the dependence is sufficiently slow that it can be neglected in
the process of integrating the Einstein field equations. The perturbed
metric will therefore carry a parametric dependence upon $v$.  

The fluid's velocity vector in the background configuration is given
by $u^\alpha = (e^{-\psi} f^{-1/2}, 0, 0, 0)$. In the perturbed
configuration it becomes $\hat{u}^\alpha = (\hat{u}^v, 0, 0, 0)$,
reflecting the fact that the tide is stationary and does not create
motion within the fluid. The time component of the vector changes by
virtue of the fact that the metric changes; we have that $\hat{u}^v =
e^{-\psi} f^{-1/2} + \delta u^v$, with $\delta u^v = \frac{1}{2}
e^{-3\psi} f^{-3/2} p_{vv}$. 

After lowering the index on $\hat{u}^\alpha$ with the perturbed
metric $g^0_{\alpha\beta} + p_{\alpha\beta}$, we find that $\hat{u}_v
=  -e^\psi f^{1/2} + \delta u_v$, $\hat{u}_r = f^{-1/2} + \delta u_r$,
and $\hat{u}_A = \delta u_A$, with 
\begin{subequations} \allowdisplaybreaks 
\label{u_pert} 
\begin{align} 
\delta u_v &= \frac{1}{2} e^{-\psi} f^{-1/2}\, p_{vv}, \\
\delta u_r &= \frac{1}{2} e^{-2\psi} f^{-3/2}\, p_{vv}, \\ 
\delta u_A &= e^{-\psi} f^{-1/2}\, p_{vA}. 
\end{align}
\end{subequations} 
These expressions are valid in the light-cone gauge. The perturbation
$\delta u_A$ can be decomposed into even-parity and odd-parity
components; the perturbations $\delta u_v$ and $\delta u_r$ are
necessarily of even parity. 

The perturbation in the energy-momentum tensor is generated by the
perturbation in $u_\alpha$, but also by a perturbation in the density
$\rho$ and pressure $p$ created by the tide; these are related by the
equation of state. We have  
\begin{eqnarray} 
\delta T_{\alpha\beta} &=& (\rho + p) \bigl( u_\alpha \delta u_\beta 
+ u_\beta \delta u_\alpha \bigr) + p\, p_{\alpha\beta} 
\nonumber \\ & & \mbox{} 
+ (\delta \rho + \delta p) u_\alpha u_\beta 
+ (\delta p) g_{\alpha\beta}, 
\end{eqnarray} 
and in the light-cone gauge this reads 
\begin{subequations} \allowdisplaybreaks 
\label{EMtensor_pert}
\begin{align} 
\delta T_{vv} &= -\rho\, p_{vv} + e^{2\psi} f\, \delta \rho, \\ 
\delta T_{vr} &= -e^\psi\, \delta \rho, \\
\delta T_{vA} &= -\rho\, p_{vA}, \\ 
\delta T_{rr} &= (\rho + p) e^{-2\psi} f^{-2}\, p_{vv} 
 + f^{-1} (\delta \rho + \delta p), \\ 
\delta T_{rA} &= e^{-\psi} f^{-1} (\rho + p)\, p_{vA}, \\
\delta T_{AB} &= p\, p_{AB} + r^2\delta p\, \Omega_{AB}. 
\end{align} 
\end{subequations} 
The perturbations $\delta T_{vA}$, $\delta T_{rA}$, and $\delta
T_{AB}$ can be decomposed into even-parity and odd-parity components;
the perturbations $\delta T_{vv}$, $\delta T_{vr}$, and 
$\delta T_{rr}$ are necessarily of even parity. 

From Eqs.~(\ref{EMtensor_pert}) we find that 
$\delta T_{\alpha\beta}$ is given by 
\begin{subequations} \allowdisplaybreaks
\label{T_even} 
\begin{align}  
\delta T_{vv} &= \sum_{lm} Q^{lm}_{vv} Y^{lm}, \\ 
\delta T_{vr} &= \sum_{lm} Q^{lm}_{vr} Y^{lm}, \\ 
\delta T_{rr} &= \sum_{lm} Q^{lm}_{rr} Y^{lm}, \\ 
\delta T_{vA} &= \sum_{lm} \bigl(Q^{lm}_v Y^{lm}_A
+ P^{lm}_v X^{lm}_A \bigr), \\
\delta T_{rA} &= \sum_{lm} \bigl(Q^{lm}_r Y^{lm}_A
+ P^{lm}_r X^{lm}_A \bigr), \\
\delta T_{AB} &= \sum_{lm} \bigl( Q^{lm}_\flat \Omega_{AB} Y^{lm}   
  + Q^{lm}_\sharp Y^{lm}_{AB} + P^{lm} X^{lm}_{AB} \bigr).  
\end{align}
\end{subequations} 
The even-parity fields are 
\begin{subequations} \allowdisplaybreaks 
\label{Q}
\begin{align} 
Q_{vv} &= -\rho\, h_{vv} + e^{2\psi} f\, \sigma, \\ 
Q_{vr} &= -e^\psi\, \sigma, \\
Q_{rr} &= (\rho + p) e^{-2\psi} f^{-2}\, h_{vv} 
 + f^{-1} (\sigma + q), \\ 
Q_{v} &= -\rho\, j_v, \\ 
Q_{r} &= e^{-\psi} f^{-1} (\rho + p)\, j_v, \\
Q_\flat &= r^2 (p\, K + q), \\ 
Q_\sharp &= r^2 p\, G, 
\end{align}
\end{subequations} 
and the perturbations in the density and pressure were also decomposed 
in spherical harmonics: 
\begin{equation} 
\delta \rho = \sum_{lm} \sigma^{lm} Y^{lm}, \qquad 
\delta p = \sum_{lm} q^{lm} Y^{lm}. 
\label{rho_and_p_pert}  
\end{equation} 
The odd-parity fields are   
\begin{subequations} \allowdisplaybreaks 
\label{P}
\begin{align} 
P_{v} &= -\rho\, h_v, \\ 
P_{r} &= e^{-\psi} f^{-1} (\rho + p)\, h_v, \\
P &= p\, h_2. 
\end{align}
\end{subequations} 

Information about $\delta\rho$ and $\delta p$, or $\sigma$ and $q$,
can be obtained from the equation of hydrostatic
equilibrium. In the perturbed spacetime the equation states that
$(\hat{\rho} + \hat{p}) \hat{a}_\alpha
+ \partial_\alpha \hat{p} = 0$, where $\hat{\rho} = \rho + \delta
\rho$ is the perturbed density, $\hat{p} = p + \delta p$ is the
perturbed pressure, and $\hat{a}_\alpha$ is the perturbed acceleration
of the fluid elements. The equation becomes 
\begin{equation} 
(\rho + p) \delta a_\alpha + (\delta \rho + \delta p) a_\alpha 
+ \partial_\alpha \delta p = 0
\label{hydrostatic} 
\end{equation} 
when expressed in terms of the perturbations $\delta \rho$, $\delta
p$, and $\delta a_\alpha$. The unperturbed acceleration has $a_r =
\frac{1}{2} e^{-2\psi} f^{-1} (e^\psi f)'$ as its only nonvanishing
component, and the perturbation has components 
\begin{subequations} \allowdisplaybreaks 
\label{a_pert} 
\begin{align} 
\delta a_v &= 0, \\ 
\delta a_r &= -\frac{1}{2} e^{-2\psi} f^{-1}\, \partial_r p_{vv} 
  + \frac{1}{2} e^{-4\psi} f^{-2} (e^{2\psi} f)'\, p_{vv}, \\
\delta a_A &= -\frac{1}{2} e^{-2\psi} f^{-1}\, \partial_A p_{vv}. 
\end{align}
\end{subequations} 
Substitution of Eqs.~(\ref{rho_and_p_pert}) and (\ref{a_pert}), as
well as $p_{vv} = \sum_{lm} h^{lm}_{vv} Y^{lm}$, into
Eq.~(\ref{hydrostatic}) reveals that 
\begin{eqnarray} 
q' &=& \frac{1}{2}(\rho+p) e^{-2\psi} f^{-1} h'_{vv} 
\nonumber \\ & & \mbox{} 
 - \frac{1}{2}(\rho+p) e^{-4\psi} f^{-2} (e^{2\psi} f)' h_{vv} 
\nonumber \\ & & \mbox{} 
- \frac{1}{2} e^{-2\psi} f^{-1} (e^{2\psi} f)' (\sigma + q) 
\label{qprime} 
\end{eqnarray}
and 
\begin{equation} 
q = \frac{1}{2} (\rho+p) e^{-2\psi} f^{-1} h_{vv}. 
\label{q}
\end{equation} 
If we next differentiate Eq.~(\ref{q}) and insert the result within
Eq.~(\ref{qprime}), we discover that 
\begin{equation} 
(\rho+p)' h_{vv} = -(e^{2\psi} f)' (\sigma + q). 
\label{sigma_plus_q} 
\end{equation} 
The last two equations allow us to express $\sigma^{lm}$ and $q^{lm}$
directly in terms of $h^{lm}_{vv}$; hydrostatic equilibrium implies
that these are not independent variables.  

\subsection{Perturbation equations: even-parity sector} 

The useful combinations of Einstein field equations are 
\begin{subequations} \allowdisplaybreaks 
\label{even_FE} 
\begin{align} 
E_1 &:= \bigl( A_{vv} - 8\pi Q_{vv} \bigr) 
 + e^\psi f \bigl( A_{vr} - 8\pi Q_{vr} \bigr) = 0, \\ 
E_2 &:= \bigl( A_{vv} - 8\pi Q_{vv} \bigr) 
 + 2e^\psi f \bigl( A_{vr} - 8\pi Q_{vr} \bigr) 
\nonumber \\ & \qquad \mbox{}
+ e^{2\psi} f^2 \bigl( A_{rr} - 8\pi Q_{rr} \bigr) = 0, \\ 
E_3 &:= \bigr( A_{rr} - 8\pi Q_{rr} \bigr) = 0, \\
E_4 &:= e^{-\psi} r E_2 + 2f \bigl( A_v - 8\pi Q_v ) 
\nonumber \\ & \qquad \mbox{}
+ 2e^\psi f^2 \bigr( A_r - 8\pi Q_r \bigr) = 0.
\end{align}
\end{subequations} 
These are a set of coupled differential equations for the
variables $h_{vv}(r)$, $j_v(r)$, $K(r)$, and $G(r)$; the remaining 
field equations are redundant by virtue of the Bianchi identities. The
explicit forms reveal that $E_1 = 0$ is a first-order differential
equation for $j_v$, $E_2 = 0$ is a first-order differential equation
for $h_{vv}$, $E_3 = 0$ is a second-order differential equation for
$K$, and $E_4 = 0$ is a first-order differential equation for $G$. 

The field equations can be manipulated to yield a decoupled equation
for the master function 
\begin{eqnarray} 
\tilde{h}_{vv} &:=& h_{vv} + e^{-\psi} \bigl( e^{2\psi} f \bigr)' j_v 
 - \frac{1}{2} r^2 f \bigl( e^{2\psi} f \bigr)' G' 
\nonumber \\ 
&=& h_{vv} + \frac{2e^\psi}{r^2} \bigl( m + 4\pi r^3 p \bigr) j_v 
\nonumber \\ & & \mbox{} 
- e^{2\psi} f  \bigl( m + 4\pi r^3 p \bigr) G'. 
\label{master_function_even} 
\end{eqnarray} 
This function is gauge invariant, and it joins smoothly with the
external version of Eq.~(\ref{even_gaugeinv1}) at $r=R$. The master
equation is  
\begin{equation} 
r^2 \tilde{h}''_{vv} + A r \tilde{h}'_{vv} - B \tilde{h}_{vv} = 0, 
\label{master_eq_even} 
\end{equation} 
where 
\begin{subequations} 
\begin{align} 
A &= \frac{2}{f} \biggl[ 1 - \frac{3m}{r} - 2\pi r^2 (\rho + 3p)
\biggr], \\
B &= \frac{1}{f} \biggl[ l(l+1) - 4\pi r^2 (\rho+p) \Bigl( 3 + 
\frac{d\rho}{dp} \Bigr) \biggr]. 
\end{align} 
\end{subequations} 
The master equation is equivalent to Eq.~(27) of
Ref.~\cite{damour-nagar:09}, in which $H := e^{-2\psi} f^{-1} 
\tilde{h}_{vv}$ is used as an alternative choice of dependent
variable.    

The master equation can be derived by the following procedure. First,
integrate the field equation $E_\sharp := A_\sharp - 8\pi Q_\sharp =
0$ and obtain $j_v = \frac{1}{2} r^2 f e^\psi G'$. This implies that
$\tilde{h}_{vv} = h_{vv}$. Second, make the substitution in the other
field equations. The result is that $E_1$ now involves $h_{vv}$, $G'$,
and $G''$; $E_2$ involves $h_{vv}$, $h'_{vv}$, $K$, $K'$, and $G'$;
$E_3$ involves $h_{vv}$, $K'$, and $K''$; and $E_4$ involves
$h_{vv}$, $K$, $K'$, $G$, and $G'$. Third, differentiate $E_2$ with
respect to $r$, and use $E_1$ to eliminate the terms in $G''$, and
$E_3$ to eliminate the terms in $K''$. The result is that $E_2'$ now
involves $h_{vv}$, $h'_{vv}$, $h''_{vv}$, $K$, $K'$, $G$, and
$G'$. Fourth, construct the linear combination $rE'_2 + a E_2 + b E_4$
and determine the functions $a$ and $b$ that eliminate all terms 
involving $K$, $K'$, $G$, $G'$. The solution is unique, and the final
result is Eq.~(\ref{master_eq_even}).   

For numerical integration it is advantageous to make the same
substitution as in Eq.~(\ref{even_gaugeinv2}), 
\begin{equation} 
\tilde{h}_{vv} = -\frac{2}{(l-1)l} r^l e_{vv}(r) \E^{(l)}_m, 
\end{equation} 
and to rewrite Eq.~(\ref{master_eq_even}) as a second-order
differential equation for $e_{vv}(r)$. This function joins smoothly
with the external version of Eq.~(\ref{even_gaugeinv4}), and 
$k_{\rm el}$ is determined by matching the values of the internal and
external functions (along with their first derivatives) at $r=R$.     

\subsection{Perturbation equations: odd-parity sector} 

The useful combinations of field equations are 
\begin{subequations} \allowdisplaybreaks 
\label{odd_FE} 
\begin{align} 
O_1 &:= (B_v - 8\pi P_v) = 0, \\ 
O_2 &:= (B_v - 8\pi P_v) + e^\psi f (B_r - 8\pi P_r) = 0. 
\end{align}
\end{subequations} 
The first is a second-order differential equation for $h_v$, while the
second is a first-order differential equation for $h_2$. 

The equation $O_1 = 0$ is fully decoupled, and the perturbation
variable $h_v$ is easily shown to be gauge invariant, as it was in
the external problem. The master  variable for the odd-parity sector
is therefore $\tilde{h}_v := h_v$, and the master equation is 
\begin{equation} 
r^2 \tilde{h}''_{v} - F r \tilde{h}'_{v} - G \tilde{h}_{v} = 0, 
\label{master_eq_odd} 
\end{equation} 
where 
\begin{subequations} 
\begin{align} 
F &= \frac{4\pi r^2}{f} (\rho + p), \\
G &= \frac{1}{f} \biggl[ l(l+1) - \frac{4m}{r} 
+ 8\pi r^2 (\rho+p) \biggr]. 
\end{align} 
\end{subequations} 
This equation is equivalent to Eq.~(31) of
Ref.~\cite{damour-nagar:09}, in which $\psi := r\tilde{h}_v' 
- 2 \tilde{h}_v$ is used as an alternative choice of dependent
variable. The function $\tilde{h}_v$ joins smoothly with the 
external version of Eq.~(\ref{odd_gaugeinv1}) at $r=R$. 

For numerical integration it is advantageous to make the same
substitution as in Eq.~(\ref{odd_gaugeinv2}), 
\begin{equation} 
\tilde{h}_{v} = \frac{2}{3(l-1)l} r^{l+1} b_{v}(r) \B^{(l)}_m, 
\end{equation} 
and to rewrite Eq.~(\ref{master_eq_odd}) as a second-order
differential equation for $b_{v}(r)$. This function joins smoothly
with the external version of Eq.~(\ref{odd_gaugeinv4}), and 
$k_{\rm mag}$ is determined by matching the values of the internal and
external functions (along with their first derivatives) at $r=R$.     

\section{Implementation for polytropes} 

The relativistic Love numbers $k_{\rm el}$ and $k_{\rm mag}$ are
determined by the numerical integration of Eqs.~(\ref{master_eq_even})
and (\ref{master_eq_odd}) and matching with the external solutions at
$r=R$. This defines a simple computational procedure that can be
implemented for any choice of equation of state. In this section we 
describe the steps that are involved when the polytropic form 
\begin{equation} 
p = K \rho^{1 + 1/n}, 
\label{EOS}
\end{equation} 
is adopted; here $K$ and the polytropic index $n$ are constants.  
We choose, however, to deviate from the procedure just outlined:
Instead of integrating the master equations for the variables 
$\tilde{h}_{vv}$ and $\tilde{h}_v$,  we integrate 
{\it the complete set of independent field equations}. This 
allows us to calculate all components of the metric perturbation, and 
matching them across $r=R$ determines, in addition to the Love
numbers, the gauge parameters $a_1$, $a_2$, $a_3$, and $\alpha$
that are automatically selected by the internal
solution.\footnote{There is no strong rationale for proceeding in this
way. The honest truth is that we became aware of 
Eq.~(\ref{master_eq_even}) only after completing the numerical
work. We derived the master equation after noticing its appearance in
Refs.~\cite{hinderer:08, damour-nagar:09} and wondering why our
formulation was more complicated than theirs.}        

\subsection{Unperturbed stellar model} 

The numerical integration of Eqs.~(\ref{Einstein}) and (\ref{TOV}) is
conveniently accomplished by introducing the dimensionless variables 
$\theta$, $\mu$, and $\xi$ defined by  
\begin{equation} 
\rho = \rho_c \theta^n, \quad 
p = p_c \theta^{n+1}, \quad 
m = m_0 \mu, \quad 
r = r_0 \xi. 
\label{variables}
\end{equation} 
Here $\rho_c := \rho(r=0)$ is the central density, and $p_c := K
\rho_c^{1+1/n}$ is the central pressure. The units of mass and radius
are chosen to be 
\begin{equation} 
m_0 := 4\pi r_0^3 \rho_c, \qquad 
r_0^2 := \frac{(n+1)p_c}{4\pi \rho_c^2}, 
\label{units}
\end{equation} 
so as to simplify the form of the field equations. 

It is useful to introduce also a ``relativistic factor'' 
\begin{equation} 
b := p_c/\rho_c, 
\label{bdef}
\end{equation} 
which determines the degree with which the stellar model is
relativistic. In terms of this we have $\rho_c = b^n/K^n$, $p_c =
b^{n+1}/K^n$, and $b$ can be used in place of $\rho_c$ to label a
stellar model, given a choice $(K,n)$ of equation of state. We also
note the relation $m_0/r_0 = (n+1) b$. We find that the units $m_0$
and $r_0$ vary with $b$ even when the equation of state is fixed. To
eliminate this dependence it is useful to define the alternative units 
\begin{equation} 
M_0 = \frac{(n+1)^{3/2}}{\sqrt{4\pi}} K^{n/2}, \qquad 
R_0 = \sqrt{\frac{n+1}{4\pi}} K^{n/2}, 
\label{alt_units}
\end{equation} 
which do not depend on $b$. We have that $m_0 = M_0 b^{(3-n)/2}$ and
$r_0 = R_0 b^{(1-n)/2}$. 

In terms of the dimensionless variables the field equations
(\ref{Einstein}) and (\ref{TOV}) become    
\begin{subequations} \allowdisplaybreaks 
\label{field_equations1} 
\begin{align} 
\frac{d\mu}{d\xi} &= \xi^2 \theta^n, \\ 
\frac{d\psi}{d\xi} &= (n+1) b \frac{\xi \theta^n(1+b\theta)}{f}, \\ 
\frac{d\theta}{d\xi} &= - \frac{(\mu + b \xi^3
  \theta^{n+1})(1+b\theta)}{\xi^2 f}, 
\end{align} 
\end{subequations} 
with $f = 1 - 2(n+1)b \mu/\xi$. The boundary conditions are
$\theta(\xi=0) = 1$, $\mu(\xi=0) = 0$, and $\psi(\xi=0) = \psi_0$. In
the limit $b \to 0$ the model becomes nonrelativistic, and the
equations for $\mu$ and $\theta$ can be combined into the well-known
Lane-Emden equation; in the limit the equation for $\psi$ becomes
irrelevant. 

The formulation of Eq.~(\ref{field_equations1}) is not optimal from a
numerical point of view. For accurate integrations it is better to use 
the variable $\nu := \mu/\xi^3$ instead of $\mu$, and $x := \ln\xi$
instead of $\xi$. The system of equations becomes 
\begin{subequations} \allowdisplaybreaks 
\label{field_equations2} 
\begin{align} 
\frac{d\nu}{dx} &= \theta^n - 3\nu, \\
\frac{d\psi}{dx} &= (n+1) b \xi^2 f^{-1} \theta^n (1+b\theta), \\
\frac{d\theta}{dx} &= -\xi^2 f^{-1} (\nu + b \theta^{n+1})(1+b\theta),   
\end{align} 
\end{subequations} 
with $f = 1 - 2(n+1) b \xi^2 \nu$. The integration begins at a large
and negative value of $x$, so that $\xi = e^x$ is small, with the
starting values  
\begin{widetext} 
\begin{subequations} \allowdisplaybreaks 
\label{starting_values} 
\begin{align} 
\nu &= \frac{1}{3} - \frac{n}{30} (1+b)(1+3b) \xi^2 
+ \frac{n}{2520} (1+b)(1+3b) \bigl[ 8n-5 + (18n-20)b + (15+30n)b^2
\bigr] \xi^4 + O(\xi^6), \\ 
\theta &= 1 - \frac{1}{6} (1+b)(1+3b) \xi^2 
+ \frac{1}{360} (1+b)(1+3b) \bigl[ 3n - 2n b + (30+15n)b^2 \bigr]
\xi^4 + O(\xi^6), \\ 
\psi &= \psi_0 + \frac{1}{2} (n+1) b (1+b) \xi^2 
- \frac{1}{24} (n+1) b (1+b) \bigl[ n - 3b + (3+3n)b^2 \bigr] 
\xi^4 + O(\xi^6). 
\end{align}
\end{subequations} 
\end{widetext} 
The integration stops at $\xi = \xi_1$, where $\theta$ goes to
zero, and $\psi_0$ is chosen so that $\psi(\xi_1) = 0$. The stellar 
mass and radius are then given by  
\begin{equation} 
M = M_0 b^{(3-n)/2} \xi_1^3 \nu(\xi_1), \qquad 
R = R_0 b^{(1-n)/2} \xi_1, 
\label{mass_radius}
\end{equation}
in the units of Eq.~(\ref{alt_units}). The compactness of the body is 
measured by $C := 2M/R = 2(n+1) b \xi_1^2 \nu(\xi_1)$; this is 
dimensionless, and therefore independent of the units $M_0$ and
$R_0$. 

\subsection{Perturbation: Even-parity sector} 

The perturbation equations (\ref{even_FE}) are simplified by involving
the background field equations (\ref{Einstein}) and (\ref{TOV}). They
are also simplified by making the substitutions of
Eqs.~(\ref{variables}), (\ref{units}), and (\ref{bdef}); we therefore
write $\rho = \rho_c \theta^n$, $p = p_c \theta^{n+1}$, $r = r_0 \xi$,
and $m = m_0 \xi^3 \nu$, where $\rho_c = (n+1)b/(4\pi r_0^2)$, 
$p_c = (n+1)b^2/(4\pi r_0^2)$, and $m_0 = (n+1)b r_0$, with $\theta$
and $\nu$ (as well as $\psi$) depending on $\xi$. Finally, we use the
fact that a term $\rho'$ in the perturbation equations can be related
to $p'$ by the equation $\rho' = (d\rho/dp) p'$, with $d\rho/dp$
determined by the equation of state.   

Another useful set of substitutions is the one displayed in
Eqs.~(\ref{even_perturbation2}), along with 
\begin{equation}
r_0^2 K = \frac{2}{(l-1)l(l+2)(l+3)} r^{l+2} e_{10}(\xi) \E^{(l)}_m,
\label{e10} 
\end{equation} 
in which we replace the original variables with the radial functions
$e_1$, $e_4$, $e_7$, and $e_{10}$. These replacements are motivated by
an analysis of the perturbation equations for small values of $r$,
which reveals that $h_{vv}$ behaves as $r^l$, $j_v$ as $r^{l+1}$, $G$
as $r^l$, and $K$ as $r^{l+2}$. The numerical factors in front of
$e_{10}$ is inserted to simplify the form of the small-$r$ expansion
of $K$, as we shall see below.

The final expression of the perturbation equations is 
\begin{widetext} 
\begin{subequations} \allowdisplaybreaks
\label{even_eqns} 
\begin{align}  
0 &= E_1 = -\xi e_4' + (l+1) e^{-\psi} f^{-1}\, e_1 
- f^{-1} A_1\, e_4, 
\\ 
0 &= E_2 = -\xi e_1' + \frac{1}{2} f^{-1} A_2\, e_1 
+ l e^\psi f^{-1} B_2\, e_4
- \frac{1}{2}(l-1)(l+2) e^{2\psi}\, e_7 
\nonumber \\ &\qquad \qquad \mbox{} 
+ \frac{1}{2(l+3)} e^{2\psi} C_2 \xi^2\, e_{10} 
- \frac{1}{(l+2)(l+3)} e^{2\psi} B_2 \xi^3\, e_{10}', 
\\
0 &= E_3 = -\xi^2 e_{10}'' + (l+2)(l+3) e^{-2\psi} f^{-2} A_3\, e_1 
- (l+2) f^{-1} B_3\, e_{10} - f^{-1} C_3 \xi\, e_{10}', 
\\
0 &= E_4 = -\xi e_{7}' 
+ \frac{l(l+1)}{2(l-1)(l+2)} e^{-2\psi} f^{-2} A_4\, e_1 
+ \frac{l}{(l-1)(l+2)} e^{-\psi} f^{-2} B_4\, e_4 
\nonumber \\ &\qquad \qquad \mbox{} 
- \frac{1}{2} l\bigl[ l+3 - 4(n+1)b\xi^2\nu \bigr] f^{-1}\, e_7 
+ \frac{l(l+1)}{2(l-1)(l+2)(l+3)} f^{-1} C_4 \xi^2\, e_{10} 
\nonumber \\ &\qquad \qquad \mbox{} 
- \frac{l(l+1)}{(l-1)(l+2)^2(l+3)} 
  \bigl[ (n+1)b(\nu+b\theta^{n+1}) \bigr] f^{-1} \xi^5\, e_{10}', 
\end{align}
\end{subequations}  
where a prime indicates differentiation with respect to $\xi$, and 
\begin{subequations} \allowdisplaybreaks 
\begin{align} 
A_1 &= l+1 - 2(n+1)b\xi^2 \bigl[(l+2)\nu + b\theta^{n+1}], \\
A_2 &= (l-2)(l+1) 
 + 2(n+1)b\xi^2 \bigl[2(l+1)\nu - \theta^n(1+b\theta) \bigr], \\
B_2 &= 1 - (n+1)b\xi^2 (\nu - b\theta^{n+1}), \\
C_2 &= l-3 + 2(n+1)b\xi^2 (\nu - b\theta^{n+1}), \\
A_3 &= n\theta^{n-1} + (4n+3) b\theta^n + 3(n+1)b^2\theta^{n+1}, \\ 
B_3 &= l+3
 - (n+1)b\xi^2 \bigl[2(l+3)\nu + \theta^n(1+b\theta) \bigr], \\
C_3 &= 2(l+3)
 - (n+1)b\xi^2 \bigl[4(l+3)\nu + \theta^n(1+b\theta) \bigr], \\
A_4 &= (l-1)(l+2)
 + 2(n+1)b\xi^2 \bigl[2\nu - \theta^n(1+b\theta) \bigr], \\
B_4 &= (l-1)(l+2)
 - (n+1)b\xi^2 \bigl[ (l^2+l-4)\nu - l(l+1) b\theta^{n+1} \bigr], \\
C_4 &= l-1
 - 2(n+1)b\xi^2 (\nu + b\theta^{n+1}).
\end{align}
\end{subequations}
\end{widetext}  
A small-$\xi$ expansion of these equations, using
Eqs.~(\ref{starting_values}), reveals that $e_1 = a_0 + O(\xi^2)$,
$e_4 = a_0 e^{-\psi_0} + O(\xi^2)$, $e_7 = a_0 e^{-2\psi_0} 
+ O(\xi^2)$, and $e_{10} = a_0 e^{-2\psi_0} (1+b)[ 3(n+1)b 
+ n ] + O(\xi^2)$, where $a_0$ is a parameter that must be determined
by matching the internal and external perturbations at the stellar
boundary.   

The perturbation equations are easily written as a first-order
dynamical system for the variables $u_1 := e_1$, $u_2 := e_4$, $u_3 =
e_7$, $u_4 := e_{10}$, and $u_5 := \xi e_{10}'$. The numerical
integration is carried out with $x := \ln\xi$ as the independent
variable, and the differential equations are integrated simultaneously
with Eqs.~(\ref{field_equations2}) to determine the unperturbed
stellar model. The integration proceeds from a large and negative
value of $x$, for which $\xi = e^x$ is small, and it stops at at $\xi
= \xi_1$ where $\theta$ goes to zero.  

The term $n\theta^{n-1}$ in $A_3$ originates from a term involving
$d\rho/dp \propto \theta^{-1}$ that multiplies $\rho \propto \theta^n$
in the field equation for $K$ (or $e_{10}$). This term diverges at the
stellar boundary when $n<1$. The singularity is integrable, however,
and it can be shown that the solution for $K(r)$ (or $e_{10}$) is
actually well-behaved at the boundary. The divergence of $A_3$
nevertheless causes issues in the numerical integration of the
perturbation equations. For this reason, the accuracy achieved for $n
< 1$ is limited compared with the accuracy obtained for $n > 1$. 

The internal perturbation must match the external perturbation at $\xi
= \xi_1$, or $r=R$, the position of the stellar boundary. The five
internal functions $e_1$, $e_4$, $e_7$, $e_{10}$, and $\xi e_{10}'$
depend on one free parameter $a_0$. The external functions, on the
other hand, depend on three gauge parameters $a_1$, $a_2$, and $a_3$,
as well as the electric-type Love number $k_{\rm el}$. The five
matching conditions determine the five parameters uniquely, including
the Love number. 

We suppose that the internal functions $u_1, \cdots, u_5$ are
determined by setting $a_0 \equiv 1$ in the numerical
integrations. The desired functions $e_1, \cdots, e_{10}$ then differ
from these by an overall multiplicative factor that we denote
$\lambda^{-1}$. We have 
\begin{subequations} \allowdisplaybreaks 
\label{ein_u} 
\begin{align} 
e^{\rm in}_1 &= \lambda^{-1} u_1, \\ 
e^{\rm in}_4 &= \lambda^{-1} u_2, \\ 
e^{\rm in}_7 &= \lambda^{-1} u_3, \\ 
e^{\rm in}_{10} &= \lambda^{-1} u_4, \\ 
\xi \frac{d e^{\rm in}_{10}}{d\xi} &= \lambda^{-1} u_5, 
\end{align}
\end{subequations} 
and the matching conditions are 
\begin{subequations} \allowdisplaybreaks
\label{even_matching1} 
\begin{align} 
e^{\rm in}_1 &= e^{\rm out}_1, \\
e^{\rm in}_4 &= e^{\rm out}_4, \\
e^{\rm in}_7 &= e^{\rm out}_7, \\
e^{\rm in}_{10} &= e^{\rm out}_{10}, \\
\xi \frac{d e^{\rm in}_{10}}{d\xi} &= 
\xi \frac{d e^{\rm out}_{10}}{d\xi}, 
\end{align}
\end{subequations}
where each side of the equation is evaluated at $\xi = \xi_1$. The
external expressions for $e_1$, $e_4$, and $e_7$ are presented in
Eqs.~(\ref{radial}) and (\ref{functions}), and these must be modified
by the gauge adjustments of Eqs.~(\ref{even_gauge_withK}). 

The function $e_{10}$ is related to $K$ by Eq.~(\ref{e10}), and the
external expression for $K$ is given by Eq.~(\ref{K}). This equation
and its derivative with respect to $r$ imply that at $\xi = \xi_1$, 
\begin{subequations} \allowdisplaybreaks 
\begin{align} 
e^{\rm out}_{10} &= 2(l+2)(l+3) C^l
  \xi_1^{-2} \bigl[ a_2 + a_3 (2M/R) \bigr], \\ 
\xi \frac{d e^{\rm out}_{10}}{d\xi} &= -2(l+2)(l+3) 
  C^l \xi_1^{-2} \bigl[ (l+2) a_2 
\nonumber \\ & \qquad \mbox{} 
+ (l+3) a_3 (2M/R) \bigr], 
\end{align}
\end{subequations} 
where $C := 2M/R$ is the compactness factor. These equations can be
solved for $a_2$ and $a_3$. Involving also the matching equations and
Eqs.~(\ref{ein_u}), we arrive at  
\begin{subequations} \allowdisplaybreaks 
\label{a2_a3} 
\begin{align} 
\lambda a_2 &= \frac{\xi_1^2}{2(l+2)(l+3) C^l} 
 \Bigl[ (l+3) u_4 + u_5 \Bigr], \\ 
\lambda a_3 &= -\frac{\xi_1^2}{2(l+2)(l+3) C^{l+1}} 
 \Bigl[ (l+2) u_4 + u_5 \Bigr].
\end{align}
\end{subequations} 
We see that the gauge parameters $a_2$ and $a_3$, rescaled by the 
unknown coefficient $\lambda$, are determined by the numerical values
obtained for $u_4$ and $u_5$.  

To solve the remaining matching equations we transfer the $a_2$ and
$a_3$ terms from the right-hand side of Eqs.~(\ref{even_gauge_withK})
to the left-hand side. Taking Eqs.~(\ref{a2_a3}) into account, we form
the combinations 
\begin{subequations} \allowdisplaybreaks 
\label{w}
\begin{align} 
w_1 &:= u_1 
 + \frac{C\xi_1^2}{2(l+2)(l+3)} \Bigl[ (l+2) u_4 + u_5 \Bigr], \\ 
w_2 &:= u_2 
 - \frac{(l+1)\xi_1^2}{2(l+2)(l+3)} \Bigl[ (l+2) u_4 + u_5 \Bigr], \\ 
w_3 &:= u_3
 - \frac{\xi_1^2}{(l+2)(l+3)} \Bigl[ (l+3) u_4 + u_5 \Bigr],
\end{align}
\end{subequations} 
which can be determined numerically. Involving now
Eqs.~(\ref{radial}), the matching conditions take the explicit form 
\begin{subequations} \allowdisplaybreaks 
\label{even_matching2} 
\begin{align} 
w_1 &= A_1 \cdot \lambda + 2 B_1 \cdot (\lambda k_{\rm el}) 
 - l C \cdot (\lambda C^{l+1} a_1), \\ 
w_2 &= A_4 \cdot \lambda 
 - 2 \frac{l+1}{l} B_4 \cdot (\lambda k_{\rm el}) 
\nonumber \\ & \qquad \mbox{} 
 + \bigl[ (l-1)(l+2) + 2C \bigr] \cdot (\lambda C^{l+1} a_1), \\ 
w_3 &= A_7 \cdot \lambda + 2 B_7 \cdot (\lambda k_{\rm el}) 
 + 2l \cdot (\lambda C^{l+1} a_1); 
\end{align}
\end{subequations} 
in these expressions the functions $A_n$ and $B_n$ are evaluated at 
$r = R$, or $2M/r = C$.    

If we define a vector $\bm{w} = (w_1, w_2, w_3)$ of numerical
quantities, and another vector $\bm{p} = (\lambda, 
\lambda k_{\rm rel}, \lambda C^{l+1} a_1)$ of unknown parameters,
these equations take the form of the matrix equation $\bm{w} 
= {\sf M}\,\bm{p}$, with a matrix $\sf M$ that is known
analytically. Solving for $\bm{p}$, the Love number is finally
determined by $k_{\rm el} = p_2/p_1$.  

\subsection{Perturbation: Odd-parity sector} 

To arrive at the final form of the perturbation equations
(\ref{odd_FE}) we follow the same steps as in the even-parity
sector. These include making the substitutions of
Eqs.~(\ref{odd_perturbation2}), to replace the original variables
$h_v$ and $h_2$ with the radial functions $b_4$ and $b_7$. 

The perturbation equations are 
\begin{subequations} \allowdisplaybreaks 
\label{odd_eqns} 
\begin{align} 
0 &= O_1 = -\xi^2 b_4'' - f^{-1} F_1 \xi\, b_4' 
+ f^{-1} G_1\, b_4, \\
0 &= O_2 = -\xi b_7' - l\, b_7 + l e^{-\psi} f^{-1}\, b_4, 
\end{align}
\end{subequations}
with
\begin{subequations} \allowdisplaybreaks 
\begin{align} 
F_1 &= 2(l+1) 
 - (n+1)b\xi^2 \bigl[ 4(l+1)\nu + \theta^n(1+b\theta) \bigr], \\ 
G_1 &= (n+1)b\xi^2 \bigl[ 2(l-1)(l+2)\nu 
 + (l+3)\theta^n(1+b\theta) \bigr]. 
\end{align}
\end{subequations} 
A small-$\xi$ expansion of these equations reveals that $b_4 
= \alpha_0 + O(\xi^2)$ and $b_7 = \alpha_0 e^{-\psi_0} + O(\xi^2)$,
where $\alpha_0$ is a parameter that must be determined by matching
the internal and external perturbations at the stellar boundary.    

The perturbation equations are easily written as a first-order
dynamical system for the variables $v_1 := b_4$, $v_2 := \xi b_4'$,
and $v_3 := b_7$. 

The internal perturbation must match the external perturbation at $\xi
= \xi_1$, or $r=R$, the position of the stellar boundary. The three
internal functions $b_4$, $\xi b_4'$, and $b_7$ depend on one free
parameter $\alpha_0$. The external functions, on the other
hand, depend on one gauge parameter $\alpha$ as well as the
magnetic-type Love number $k_{\rm mag}$. The three matching conditions 
determine the three parameters uniquely, including the Love number.  

We suppose that the perturbation equations for $v_1$, $v_2$, and $v_3$
are integrated with $\alpha_0 \equiv 1$. The desired internal
functions $b_4$ and $b_7$ are then given by 
\begin{subequations} \allowdisplaybreaks 
\label{bin_v}
\begin{align} 
b_4^{\rm in} &= \lambda^{-1} v_1, \\ 
\xi \frac{d b_4^{\rm in}}{d\xi} &= \lambda^{-1} v_2, \\ 
b_7^{\rm in} &= \lambda^{-1} v_3, 
\end{align} 
\end{subequations} 
where $\lambda$ is an unknown constant. The matching conditions are 
\begin{subequations} \allowdisplaybreaks
\label{odd_matching1}  
\begin{align} 
b_4^{\rm in} &= b_4^{\rm out}, \\ 
\xi \frac{d b_4^{\rm in}}{d\xi} &= 
  \xi \frac{d b_4^{\rm out}}{d\xi}, \\ 
b_7^{\rm in} &= b_7^{\rm out}, 
\end{align} 
\end{subequations} 
where each side of the equation is evaluated at $\xi = \xi_1$. The
external expressions for $b_4$ and $b_7$ are presented in
Eqs.~(\ref{radial}), together with the gauge adjustment of
Eq.~(\ref{odd_gauge}). We observe that $b_4$ is gauge-invariant, and
that the purpose of the matching equation for $b_7$ is to determine
the (uninteresting) gauge parameter $\alpha$.  

We focus on the two equations involving $b_4$. Using
Eqs.~(\ref{radial}), we find that the explicit form of the
matching conditions is
\begin{subequations} \allowdisplaybreaks 
\label{odd_matching2} 
\begin{align} 
v_1 &= A_4 \cdot \lambda 
  - 2 \frac{l+1}{l} B_4 \cdot (\lambda k_{\rm mag}), \\ 
v_2 &= -C A_4' \cdot \lambda  
  + 2 \frac{l+1}{l} \bigl[ C B_4' + (2l+1) B_4 \bigr] 
  \cdot (\lambda k_{\rm mag}).
\end{align}
\end{subequations}  
In these expressions the functions $A_4$, $A_4' := dA_4/dz$, $B_4$,
and $B_4' := dB_4/dz$ are evaluated at $z := 2M/r = C$.        

If we define a vector $\bm{v} = (v_1, v_2)$ of numerical
quantities, and another vector $\bm{p} = (\lambda, 
\lambda k_{\rm mag})$ of unknown parameters, these equations take the
form of the matrix equation $\bm{v} = {\sf M}\,\bm{p}$, with a matrix
$\sf M$ that is known analytically. Solving for $\bm{p}$, the Love
number is finally determined by $k_{\rm mag} = p_2/p_1$.  

To evaluate the derivatives of $A_4$ and $B_4$ with respect to $z$ we
use the well-known property of hypergeometric functions that $(d/dz)
F(a,b;c;z) = (ab/c) F(a+1,b+1;c+1;z)$. 

\section{Numerical results} 

The computations presented in this work were generated with two
independent codes, one written by each author. Consistency between our 
results provides evidence that each set of computations were carried
our correctly, and the comparison allows us to estimate the numerical
accuracy of our results.  

The background spacetime is constructed by solving the Einstein field
equations for a spherical matter configuration with a polytropic
equation of state. The equations were formulated in Sec.~V A, and the  
system of equations (\ref{field_equations2}) is integrated
numerically for selected values of the polytropic index $n$. The
integration begins at a large and negative value of 
the radial variable $x = \ln\xi$, using the starting values listed in 
Eqs.~(\ref{starting_values}). It proceeds until $\theta$ changes sign
at the stellar boundary, $x = x_1$. In the first code, the integration
is performed using the Bulirsh-Stoer method as implemented in the 
{\it Numerical Recipes} routine {\tt  bsstep}, which is embedded
within {\tt odeint}; we use the Second Edition of {\it Numerical
Recipes} \cite{numerical-recipes:92}, and the code is written in
C++. In the second code, the integration is performed using the
embedded Runge-Kutta Prince-Dormand method as implemented in the
{\it GNU Scientific Library} routine {\tt rk8pd}, which is embedded
within {\tt odeiv}; we use version 1.9 of the libraries \cite{GSL},
and the code is written in C. In each code all floating-point 
operations are carried out with double precision. The accuracy of the
integration is determined by the integrator's tolerance $\epsilon$ and
the errors of order $\xi^6$ that are incorporated in the starting
values. As Eqs.~(\ref{field_equations2}) are exceptionally well
conditioned toward numerical integration, a high degree of accuracy
can easily be achieved. We estimate that our stellar configurations
are computed accurately to at least twelve significant digits.  

The stellar boundary is identified with the help of a bisection search 
for the solution to $\theta(x) = 0$. In the first code this is carried
out with the {\it Numerical Recipe} routine {\tt zbrent}; the search
is loosely bracketed between the values $x_0 < x_1$ (where $\theta$ is
positive) and $x_2 > x_1$ (where $\theta$ is negative). In the second
code this is carried out with the {\it GNU Scientific Library} routine
{\tt brent}, using a similar bracketing method. The search is
carried out with high accuracy, again of the order of twelve
significant digits.    

The even-parity perturbation equations (\ref{even_eqns}) are
next integrated for selected values of $n$ and $l$, simultaneously
with the background field equations (\ref{field_equations2}). Once
more the integration begins at a large and negative value of $x$,
using the starting values derived in Sec.~V B, and it
proceeds up to $x = x_1$. In the first code we continue to use 
{\tt bsstep} and  {\tt odeint}, and the caption of Table II discusses
the accuracy of these integrations. In the second code we continue to
use {\tt rk8pd} and {\tt odeiv}; the tolerance of the integrator is
set uniformly to $\epsilon = 1.0\e{-12}$, and all integrations begin
at $x = -10.0$. Each code returns the values of $u_1$, $u_2$, $u_3$, 
$u_4$, and $u_5$ at the stellar boundary.      

The odd-parity equations (\ref{odd_eqns}) are integrated in
exactly the same way. Here the codes return the values of $v_1$,
$v_2$, and $v_3$ at the stellar boundary. 

The matching problem of Eqs.~(\ref{even_matching2}) requires the
numerical solution of the matrix equation $\bm{w} = {\sf M}\, \bm{p}$, 
where $\bm{w}$ is constructed from the perturbations, $\sf M$ is 
known analytically, and $\bm{p}$ is the vector of unknown parameters,
which include the electric-type Love number $k_{\rm el}$. In the first
code the system of equations is solved by performing an LU
decomposition of the matrix $\sf M$, and this is handled by the 
{\it Numerical Recipes} routines {\tt ludcmp} and 
{\tt lubksb}. In the second code the LU decomposition is handled by
the {\it GNU Scientific Library} routines 
{\tt gsl\_linalg\_LU\_decomp} 
and {\tt gsl\_linalg\_LU\_solve}. In view of the small number
of equations involved (three), this task is essentially carried out at
machine precision. The final output is $k_{\rm el}$.   

The matching problem of Eqs.~(\ref{odd_matching2}) is handled in 
exactly the same way. Here the final output is the magnetic-type Love
number $k_{\rm mag}$. 

Our results are presented in the figures displayed in Sec.~I and in
the tables provided in the Appendix. The electric-type and
magnetic-type Love numbers are computed for selected values of $n$ and
$l$, as functions of the relativistic parameter $b := p_c/\rho_c$ and
the compactness $C := 2M/R$. The allowed interval begins at
$b=0$ and $C=0$, where the equations reduce to their Newtonian limit,
and ends at $b = b_{\rm max}$ and $C = C_{\rm max}$, where the stellar
configuration achieves its maximum mass.  Each table caption discusses
the estimated accuracy of our results. Overall we claim an approximate
accuracy of nine significant digits for the Love numbers (with some
exceptions, as detailed in the table captions).    

\begin{table}[H]
\caption{Integration errors for even-parity perturbations. For each
  selected value of $n$ the first row shows the value of $\epsilon$,
  the integrator's tolerance. When $\epsilon = 1.0\e{-12}$ the
  integrations are started at $x = -7.0$, so that the errors in the
  starting values are of the order of $1.0\e{-12}$. When $\epsilon >
  1.0\e{-12}$ the integrations are started at $x = -6.5$, so that the
  errors in the starting values are of the order of $1.0\e{-11}$. For
  the odd-parity equations the tolerance of the integrator is set
  uniformly to $\epsilon = 1.0\e{-12}$, and all integrations begin at
  $x = -7.0$. The second column shows $\delta$, an intrinsic measure
  of the accuracy of our results. This is defined as $\delta 
  := |\nu_{\rm model} - \nu_{\rm pert}|/\nu_{\rm model}$, where
  $\nu_{\rm model}$ is the value of $\nu$ at the stellar boundary $\xi
  = \xi_1$ as determined with exquisite precision by integrating  the
  stellar-model equations only, while $\nu_{\rm pert}$ is the value as
  determined by also integrating the perturbation equations. The least 
  accurate determinations are for small values of $b$; the accuracy
  typically improves by two orders of magnitude at larger values of
  $b$. For reasons that were explained at the end of Sec.~IV D, when 
  $n<1$ the accuracy that can be achieved for the even-parity
  perturbations is more limited than what is achieved for $\nu$; for
  these cases $\delta$ gives an overestimate of the true accuracy. For
  $n > 1$, and for the odd-parity perturbations, $\delta$ should be an 
  accurate measure of our accuracy.}   
\begin{ruledtabular} 
\begin{tabular}{lllll}
                   & $l = 2$        & $l = 3$        & $l = 4$         & $l = 5$ \\
\hline
$n = 0.50 $ & $\epsilon = 1.0\e{-10}$ & $\epsilon = 1.0\e{-10}$ &
$\epsilon = 1.0\e{-10}$ & $\epsilon = 1.0\e{-10}$ \\
                    & $\delta < 1.2\e{-10}$ & $\delta < 1.2\e{-10}$ &
$\delta < 1.2\e{-10}$& $\delta < 1.2\e{-10}$ \\ 
$n = 0.75 $ & $\epsilon = 3.0\e{-11}$ & $\epsilon = 3.0\e{-11}$ &
$\epsilon = 3.0\e{-11}$ & $\epsilon = 3.0\e{-11}$ \\
                    & $\delta < 8.6\e{-11}$ & $\delta < 6.6\e{-11}$ &
$\delta < 8.6\e{-11}$& $\delta < 8.6\e{-11}$ \\ 
$n = 1.00 $ & $\epsilon = 1.0\e{-12}$ & $\epsilon = 3.0\e{-11}$ &
$\epsilon = 3.0\e{-11}$ & $\epsilon = 3.0\e{-11}$ \\
                    & $\delta < 1.6\e{-09}$ & $\delta < 1.7\e{-09}$ &
$\delta < 2.7\e{-10}$& $\delta < 4.0\e{-11}$ \\ 
$n = 1.25 $ & $\epsilon = 1.0\e{-12}$ & $\epsilon = 3.0\e{-11}$ &
$\epsilon = 3.0\e{-11}$ & $\epsilon = 3.0\e{-11}$ \\
                    & $\delta < 9.5\e{-11}$ & $\delta < 9.6\e{-11}$ &
$\delta < 9.5\e{-11}$& $\delta < 9.5\e{-11}$ \\ 
$n = 1.50 $ & $\epsilon = 1.0\e{-12}$ & $\epsilon = 3.0\e{-11}$ &
$\epsilon = 3.0\e{-11}$ & $\epsilon = 3.0\e{-11}$ \\
                    & $\delta < 7.2\e{-11}$ & $\delta < 7.2\e{-11}$ &
$\delta < 7.2\e{-11}$& $\delta < 7.2\e{-11}$ \\ 
$n = 1.75 $ & $\epsilon = 1.0\e{-12}$ & $\epsilon = 1.0\e{-12}$ &
$\epsilon = 7.0\e{-11}$ & $\epsilon = 7.0\e{-11}$ \\
                    & $\delta < 9.2\e{-11}$ & $\delta < 9.2\e{-11}$ &
$\delta < 9.2\e{-11}$& $\delta < 9.2\e{-11}$ \\ 
$n = 2.00 $ & $\epsilon = 1.0\e{-12}$ & $\epsilon = 1.0\e{-12}$ &
$\epsilon = 1.0\e{-12}$ & $\epsilon = 3.0\e{-11}$ \\
                    & $\delta < 2.4\e{-12}$ & $\delta < 2.4\e{-12}$ &
$\delta < 2.4\e{-12}$& $\delta < 2.4\e{-12}$
\end{tabular}
\end{ruledtabular} 
\end{table}

\begin{acknowledgments} 
This work was supported by the Natural Sciences and Engineering
Research Council of Canada. We thank Thibault Damour, Eanna Flanagan,
Tanja Hinderer, Scott Hughes, Ryan Lang, and Alessandro Nagar for
useful discussions. In addition, we thank an anonymous referee who
helped us make substantial improvements to the paper. EP would also
like to thank the staff of the Canadian Institute for Theoretical
Astrophysics for their kind hospitality during the time of his
research leave; a large of portion of this work was completed during
this time.    
\end{acknowledgments} 

\appendix*
\section{Tables of relativistic Love numbers} 

\begingroup 
\squeezetable 

\begin{table}[H]
\caption{Love numbers for $n = 0.50$ and $l=2$. Integration of the
  Newtonian Clairaut equation (for $b=0$) returns $k_{\rm el} 
  = 4.491539995415\e{-01}$. This provides evidence that our results
  for the electric-type Love numbers are accurate to five significant
  digits. We believe that our results for the magnetic-type Love
  numbers are accurate to nine significant digits.}  
\begin{ruledtabular} 
\begin{tabular}{llll}
$ b$ & $2M/R$ & $k_{\rm el}$ & $k_{\rm mag}$ \\ 
\hline 
0.0000000000 & 0.0000000000 & 4.4915295584e-01 & 0.0000000000e+00 \\ 
0.0162962963 & 0.0627859865 & 3.6857599573e-01 & 1.6896831556e-03 \\ 
0.0651851852 & 0.2085406132 & 2.2117103643e-01 & 4.1596525372e-03 \\ 
0.1466666667 & 0.3636165454 & 1.1528493484e-01 & 4.8522931593e-03 \\ 
0.2607407407 & 0.4883414066 & 6.0195686393e-02 & 4.2583493395e-03 \\ 
0.4074074074 & 0.5772867923 & 3.3825660571e-02 & 3.3759103748e-03 \\ 
0.5866666667 & 0.6379537736 & 2.0879830949e-02 & 2.6271666539e-03 \\ 
0.7985185185 & 0.6789539591 & 1.4107892752e-02 & 2.0809271632e-03 \\ 
1.0429629630 & 0.7068171264 & 1.0311605798e-02 & 1.7004161103e-03 \\ 
1.3200000000 & 0.7259502382 & 8.0453742292e-03 & 1.4372173622e-03
\end{tabular}
\end{ruledtabular} 
\end{table}

\begin{table}[H]
\caption{Love numbers for $n = 0.75$ and $l=2$. Integration of the
  Newtonian Clairaut equation (for $b=0$)  returns $k_{\rm el} 
  = 3.434291771770\e{-01}$. This provides evidence that our results
  for the electric-type Love numbers are accurate to nine significant
  digits. We believe that our results for the magnetic-type Love
  numbers are also accurate to nine significant digits.}  
\begin{ruledtabular} 
\begin{tabular}{llll}
$ b$ & $2M/R$ & $k_{\rm el}$ & $k_{\rm mag}$ \\ 
\hline
0.0000000000 & 0.0000000000 & 3.4342917761e-01 & 0.0000000000e+00 \\ 
0.0092592593 & 0.0363293144 & 3.0528456672e-01 & 8.4958217044e-04 \\ 
0.0370370370 & 0.1294393332 & 2.2114677984e-01 & 2.5151747439e-03 \\ 
0.0833333333 & 0.2455878442 & 1.4055702918e-01 & 3.6455944426e-03 \\ 
0.1481481481 & 0.3564603549 & 8.4890119064e-02 & 3.8700558106e-03 \\ 
0.2314814815 & 0.4485200887 & 5.1708383719e-02 & 3.5282781348e-03 \\ 
0.3333333333 & 0.5194393410 & 3.2857454025e-02 & 3.0026963633e-03 \\ 
0.4537037037 & 0.5719839827 & 2.2091374455e-02 & 2.4979305215e-03 \\ 
0.5925925926 & 0.6101589410 & 1.5755967215e-02 & 2.0827696087e-03 \\ 
0.7500000000 & 0.6376107260 & 1.1882367812e-02 & 1.7629062677e-03
\end{tabular}
\end{ruledtabular} 
\end{table}

\begin{table}[H]
\caption{Love numbers for $n = 1.00$ and $l=2$. Integration of the
  Newtonian Clairaut equation (for $b=0$)  returns $k_{\rm el} 
  = 2.599088771480\e{-01}$. This provides evidence that our results
  for the electric-type Love numbers are accurate to nine significant
  digits. We believe that our results for the magnetic-type Love
  numbers are also accurate to nine significant digits.} 
\begin{ruledtabular} 
\begin{tabular}{llll}
$ b$ & $2M/R$ & $k_{\rm el}$ & $k_{\rm mag}$\\ 
\hline
0.0000000000 & 0.0000000000 & 2.5990887732e-01 & 0.0000000000e+00 \\ 
0.0054320988 & 0.0211887760 & 2.4198937486e-01 & 4.1832500500e-04 \\ 
0.0217283951 & 0.0788326459 & 1.9761362790e-01 & 1.3874544444e-03 \\ 
0.0488888889 & 0.1586178173 & 1.4594601117e-01 & 2.3418562643e-03 \\ 
0.0869135802 & 0.2449940757 & 1.0135201470e-01 & 2.9092025152e-03 \\ 
0.1358024691 & 0.3264977638 & 6.8656738911e-02 & 3.0470778168e-03 \\ 
0.1955555556 & 0.3971100356 & 4.6672564713e-02 & 2.8932146678e-03 \\ 
0.2661728395 & 0.4550360296 & 3.2438798694e-02 & 2.6030597661e-03 \\ 
0.3476543210 & 0.5008905693 & 2.3293063321e-02 & 2.2824210048e-03 \\ 
0.4400000000 & 0.5363092473 & 1.7360105151e-02 & 1.9854445481e-03
\end{tabular}
\end{ruledtabular} 
\end{table}

\begin{table}[H]
\caption{Love numbers for $n = 1.25$ and $l=2$. Integration of the
  Newtonian Clairaut equation (for  $b=0$)  returns $k_{\rm el} 
  = 1.943393766752\e{-01}$. This provides evidence that our results
  for the electric-type Love numbers are accurate to nine
  significant digits. We believe that our results for the magnetic-type Love
  numbers are also accurate to nine significant digits.} 
\begin{ruledtabular} 
\begin{tabular}{llll}
$ b$ & $2M/R$ & $k_{\rm el}$ & $k_{\rm mag}$ \\ 
\hline
0.0000000000 & 0.0000000000 & 1.9433937665e-01 & 0.0000000000e+00 \\ 
0.0037037037 & 0.0140910881 & 1.8487046323e-01 & 2.2908538860e-04 \\ 
0.0148148148 & 0.0535218477 & 1.6007564892e-01 & 8.0226880316e-04 \\ 
0.0333333333 & 0.1109782733 & 1.2818558203e-01 & 1.4646634398e-03 \\ 
0.0592592593 & 0.1774670027 & 9.7029969481e-02 & 1.9894582440e-03 \\ 
0.0925925926 & 0.2449789236 & 7.1049247827e-02 & 2.2762232723e-03 \\ 
0.1333333333 & 0.3079051472 & 5.1376476586e-02 & 2.3393293198e-03 \\ 
0.1814814815 & 0.3631764781 & 3.7288438980e-02 & 2.2472291456e-03 \\ 
0.2370370370 & 0.4096934360 & 2.7477497915e-02 & 2.0722703848e-03 \\ 
0.3000000000 & 0.4475972736 & 2.0708768325e-02 & 1.8682931012e-03
\end{tabular}
\end{ruledtabular} 
\end{table}

\begin{table}[H]
\caption{Love numbers for $n = 1.50$ and $l=2$. Integration of the
  Newtonian Clairaut equation (for $b=0$)  returns $k_{\rm el} 
  = 1.432787706403\e{-01}$. This provides evidence that our results
  for the electric-type Love numbers are accurate to nine
  significant digits. We believe that our results for the magnetic-type Love
  numbers are also accurate to nine significant digits.}  
\begin{ruledtabular} 
\begin{tabular}{llll}
$ b$ & $2M/R$ & $k_{\rm el}$ & $k_{\rm mag}$ \\ 
\hline
0.0000000000 & 0.0000000000 & 1.4327877058e-01 & 0.0000000000e+00 \\ 
0.0025925926 & 0.0095061309 & 1.3824519472e-01 & 1.2516879778e-04 \\ 
0.0103703704 & 0.0366144717 & 1.2455723004e-01 & 4.5461861960e-04 \\ 
0.0233333333 & 0.0775387088 & 1.0568541028e-01 & 8.7668437452e-04 \\ 
0.0414814815 & 0.1272141307 & 8.5491826720e-02 & 1.2719049787e-03 \\ 
0.0648148148 & 0.1805262693 & 6.6864122654e-02 & 1.5603334794e-03 \\ 
0.0933333333 & 0.2332124569 & 5.1266906792e-02 & 1.7154390472e-03 \\ 
0.1270370370 & 0.2822627839 & 3.9011593697e-02 & 1.7510464094e-03 \\ 
0.1659259259 & 0.3259042474 & 2.9758544257e-02 & 1.6998078860e-03 \\ 
0.2100000000 & 0.3633567807 & 2.2929142045e-02 & 1.5963449703e-03
\end{tabular}
\end{ruledtabular} 
\end{table}

\begin{table}[H]
\caption{Love numbers for $n = 1.75$ and $l=2$. Integration of the
  Newtonian Clairaut equation (for $b=0$)  returns $k_{\rm el} 
  = 1.039154459896\e{-01}$. This provides evidence that our results
  for the electric-type Love numbers are accurate to nine
  significant digits. We believe that our results for the magnetic-type Love
  numbers are also accurate to nine significant digits.}  
\begin{ruledtabular} 
\begin{tabular}{llll}
$ b$ & $2M/R$ & $k_{\rm el}$ & $k_{\rm mag}$ \\ 
\hline
0.0000000000 & 0.0000000000 & 1.0391544596e-01 & 0.0000000000e+00 \\ 
0.0018518519 & 0.0064743298 & 1.0123582505e-01 & 6.7882541528e-05 \\ 
0.0074074074 & 0.0251788386 & 9.3755924816e-02 & 2.5279292539e-04 \\ 
0.0166666667 & 0.0541259986 & 8.2923108986e-02 & 5.0667193661e-04 \\ 
0.0296296296 & 0.0904962838 & 7.0531274857e-02 & 7.7147592064e-04 \\ 
0.0462962963 & 0.1311832131 & 5.8177787311e-02 & 9.9862752282e-04 \\ 
0.0666666667 & 0.1732771212 & 4.6952247447e-02 & 1.1599057338e-03 \\ 
0.0907407407 & 0.2143825635 & 3.7393783155e-02 & 1.2480829694e-03 \\ 
0.1185185185 & 0.2527489948 & 2.9616069949e-02 & 1.2710091094e-03 \\ 
0.1500000000 & 0.2872548584 & 2.3478510254e-02 & 1.2440214762e-03
\end{tabular}
\end{ruledtabular} 
\end{table}

\begin{table}[H]
\caption{Love numbers for $n = 2.00$ and $l=2$. Integration of the
  Newtonian Clairaut equation (for $b=0$)  returns $k_{\rm el} 
  = 7.393839192094\e{-02}$. This provides evidence that our results
  for the electric-type Love numbers are accurate to nine
  significant digits. We believe that our results for the magnetic-type Love
  numbers are also accurate to nine significant digits.} 
\begin{ruledtabular} 
\begin{tabular}{llll}
$ b$ & $2M/R$ & $k_{\rm el}$ & $k_{\rm mag}$ \\ 
\hline
0.0000000000 & 0.0000000000 & 7.3938391925e-02 & 0.0000000000e+00 \\ 
0.0013580247 & 0.0044806121 & 7.2500928560e-02 & 3.6722504616e-05 \\ 
0.0054320988 & 0.0175403971 & 6.8415857342e-02 & 1.3908592972e-04 \\ 
0.0122222222 & 0.0381005822 & 6.2293615431e-02 & 2.8628982295e-04 \\ 
0.0217283951 & 0.0645673693 & 5.4948705679e-02 & 4.5110089827e-04 \\ 
0.0339506173 & 0.0950758472 & 4.7195018104e-02 & 6.0738825840e-04 \\ 
0.0488888889 & 0.1277344111 & 3.9692062306e-02 & 7.3574977114e-04 \\ 
0.0665432099 & 0.1608206440 & 3.2876266347e-02 & 8.2584357166e-04 \\ 
0.0869135802 & 0.1929044035 & 2.6967366952e-02 & 8.7572897640e-04 \\ 
0.1100000000 & 0.2228982181 & 2.2017664632e-02 & 8.8947977626e-04
\end{tabular}
\end{ruledtabular} 
\end{table}

\begin{table}[H]
\caption{Love numbers for $n = 0.50$ and $l=3$. Integration of the
  Newtonian Clairaut equation (for $b=0$) returns $k_{\rm el} 
  = 2.033844048605\e{-01}$. This provides evidence that our results
  for the electric-type Love numbers are accurate to five significant
  digits. We believe that our results for the magnetic-type Love
  numbers are accurate to nine significant digits.} 
\begin{ruledtabular} 
\begin{tabular}{llll}
$ b$ & $2M/R$ & $k_{\rm el}$ & $k_{\rm mag}$ \\ 
\hline 
0.0000000000 & 0.0000000000 & 2.0338399420e-01 & 0.0000000000e+00 \\ 
0.0162962963 & 0.0627859865 & 1.5613095764e-01 & 7.6806695868e-04 \\ 
0.0651851852 & 0.2085406132 & 7.9298498872e-02 & 1.5334802180e-03 \\ 
0.1466666667 & 0.3636165454 & 3.3876635518e-02 & 1.4053484595e-03 \\ 
0.2607407407 & 0.4883414066 & 1.4803140981e-02 & 1.0040207606e-03 \\ 
0.4074074074 & 0.5772867923 & 7.2690559784e-03 & 6.8628867282e-04 \\ 
0.5866666667 & 0.6379537736 & 4.0970043659e-03 & 4.8495681685e-04 \\ 
0.7985185185 & 0.6789539591 & 2.6182028430e-03 & 3.6226354828e-04 \\ 
1.0429629630 & 0.7068171264 & 1.8555610194e-03 & 2.8623307831e-04 \\ 
1.3200000000 & 0.7259502382 & 1.4266248200e-03 & 2.3757226341e-04
\end{tabular}
\end{ruledtabular} 
\end{table}

\begin{table}[H]
\caption{Love numbers for $n = 0.75$ and $l=3$. Integration of the
  Newtonian Clairaut equation (for $b=0$) returns $k_{\rm el} 
  = 1.479565910794\e{-01}$, and this value was copied in the first row
  of the Table. [We were not able to accurately compute the
  electric-type Love number for $b=0$ for these specific values of $n$
  and $l$. The reason has to do with the fact that for these values,
  $\xi e_{10}' = O(\xi^4)$ instead of being of order $\xi^2$ near $\xi
  = 0$; the integrator then has difficulty moving out of the
  small-$\xi$ region and the number of steps required exceeds the set
  limit.] We believe that our results for the electric-type Love
  numbers are accurate to nine significant digits, and that our results
  for the magnetic-type Love numbers are also accurate to nine
  significant digits.}    
\begin{ruledtabular} 
\begin{tabular}{llll}
$ b$ & $2M/R$ & $k_{\rm el}$ & $k_{\rm mag}$ \\ 
\hline 
0.0000000000 & 0.0000000000 & 1.4795659108e-01 & 0.0000000000e+00 \\
0.0092592593 & 0.0363293144 & 1.2656912474e-01 & 3.7606006609e-04 \\ 
0.0370370370 & 0.1294393332 & 8.2729118342e-02 & 9.7442657049e-04 \\ 
0.0833333333 & 0.2455878442 & 4.5802954407e-02 & 1.1847104807e-03 \\ 
0.1481481481 & 0.3564603549 & 2.3980661301e-02 & 1.0533665372e-03 \\ 
0.2314814815 & 0.4485200887 & 1.2870072782e-02 & 8.2451036345e-04 \\ 
0.3333333333 & 0.5194393410 & 7.3958196272e-03 & 6.2355209034e-04 \\ 
0.4537037037 & 0.5719839827 & 4.6219154497e-03 & 4.7654009722e-04 \\ 
0.5925925926 & 0.6101589410 & 3.1388009813e-03 & 3.7510397112e-04 \\ 
0.7500000000 & 0.6376107260 & 2.2971603620e-03 & 3.0589573877e-04
\end{tabular}
\end{ruledtabular} 
\end{table}

\begin{table}[H]
\caption{Love numbers for $n = 1.00$ and $l=3$. Integration of the
  Newtonian Clairaut equation (for $b=0$) returns $k_{\rm el} 
  = 1.064540469774\e{-01}$. This provides evidence that our results
  for the electric-type Love numbers are accurate to nine
  significant digits. We believe that our results for the magnetic-type Love
  numbers are also accurate to nine significant digits.} 
\begin{ruledtabular} 
\begin{tabular}{llll}
$ b$ & $2M/R$ & $k_{\rm el}$ & $k_{\rm mag}$ \\ 
\hline 
0.0000000000 & 0.0000000000 & 1.0645404707e-01 & 0.0000000000e+00 \\ 
0.0054320988 & 0.0211887760 & 9.6920315090e-02 & 1.7712436105e-04 \\ 
0.0217283951 & 0.0788326459 & 7.4354157385e-02 & 5.4046973625e-04 \\ 
0.0488888889 & 0.1586178173 & 5.0164622016e-02 & 8.0927197289e-04 \\ 
0.0869135802 & 0.2449940757 & 3.1412814100e-02 & 8.7805794150e-04 \\ 
0.1358024691 & 0.3264977638 & 1.9197799362e-02 & 8.0536598779e-04 \\ 
0.1955555556 & 0.3971100356 & 1.1894296947e-02 & 6.7961132670e-04 \\ 
0.2661728395 & 0.4550360296 & 7.6520308738e-03 & 5.5466172884e-04 \\ 
0.3476543210 & 0.5008905693 & 5.1742095921e-03 & 4.5069370183e-04 \\ 
0.4400000000 & 0.5363092473 & 3.6915952107e-03 & 3.7043176559e-04
\end{tabular}
\end{ruledtabular} 
\end{table}

\begin{table}[H]
\caption{Love numbers for $n = 1.25$ and $l=3$. Integration of the
  Newtonian Clairaut equation (for $b=0$) returns $k_{\rm el} 
  = 7.558993098406\e{-02}$. This provides evidence that our results
  for the electric-type Love numbers are accurate to eight significant
  digits. We believe that our results for the magnetic-type Love
  numbers are accurate to nine significant digits.} 
\begin{ruledtabular} 
\begin{tabular}{llll}
$ b$ & $2M/R$ & $k_{\rm el}$ & $k_{\rm mag}$ \\ 
\hline 
0.0000000000 & 0.0000000000 & 7.5589930713e-02 & 0.0000000000e+00 \\
0.0037037037 & 0.0140910881 & 7.0841000307e-02 & 9.1484438895e-05 \\ 
0.0148148148 & 0.0535218477 & 5.8788456530e-02 & 3.0220575213e-04 \\ 
0.0333333333 & 0.1109782733 & 4.4167684362e-02 & 5.0553940166e-04 \\ 
0.0592592593 & 0.1774670027 & 3.0965107128e-02 & 6.1848609429e-04 \\ 
0.0925925926 & 0.2449789236 & 2.0910331154e-02 & 6.3402270724e-04 \\ 
0.1333333333 & 0.3079051472 & 1.3985888466e-02 & 5.8644352242e-04 \\ 
0.1814814815 & 0.3631764781 & 9.4660656773e-03 & 5.1264600426e-04 \\ 
0.2370370370 & 0.4096934360 & 6.5780656202e-03 & 4.3641911298e-04 \\ 
0.3000000000 & 0.4475972736 & 4.7331767581e-03 & 3.6881598982e-04
\end{tabular}
\end{ruledtabular} 
\end{table}

\begin{table}[H]
\caption{Love numbers for $n = 1.50$ and $l=3$. Integration of the
  Newtonian Clairaut equation (for $b=0$) returns $k_{\rm el} 
  = 5.284852444148\e{-02}$. This provides evidence that our results
  for the electric-type Love numbers are accurate to eight significant
  digits. We believe that our results for the magnetic-type Love
  numbers are accurate to nine significant digits.} 
\begin{ruledtabular} 
\begin{tabular}{llll}
$ b$ & $2M/R$ & $k_{\rm el}$ & $k_{\rm mag}$ \\ 
\hline 
0.0000000000 & 0.0000000000 & 5.2848524127e-02 & 0.0000000000e+00 \\
0.0025925926 & 0.0095061309 & 5.0478328434e-02 & 4.6818602757e-05 \\ 
0.0103703704 & 0.0366144717 & 4.4172676367e-02 & 1.6314544473e-04 \\ 
0.0233333333 & 0.0775387088 & 3.5833784527e-02 & 2.9516806482e-04 \\ 
0.0414814815 & 0.1272141307 & 2.7409928071e-02 & 3.9553177536e-04 \\ 
0.0648148148 & 0.1805262693 & 2.0154523547e-02 & 4.4448482790e-04 \\ 
0.0933333333 & 0.2332124569 & 1.4516418814e-02 & 4.4702358096e-04 \\ 
0.1270370370 & 0.2822627839 & 1.0411153801e-02 & 4.1916508131e-04 \\ 
0.1659259259 & 0.3259042474 & 7.5326749926e-03 & 3.7680447176e-04 \\ 
0.2100000000 & 0.3633567807 & 5.5503018177e-03 & 3.3107508747e-04
\end{tabular}
\end{ruledtabular} 
\end{table}

\begin{table}[H]
\caption{Love numbers for $n = 1.75$ and $l=3$. Integration of the
  Newtonian Clairaut equation (for $b=0$) returns $k_{\rm el} 
  = 3.628620386492\e{-02}$. This provides evidence that our results
  for the electric-type Love numbers are accurate to nine significant
  digits. We believe that our results for the magnetic-type Love
  numbers are also accurate to nine significant digits.} 
\begin{ruledtabular} 
\begin{tabular}{llll}
$ b$ & $2M/R$ & $k_{\rm el}$ & $k_{\rm mag}$ \\ 
\hline 
0.0000000000 & 0.0000000000 & 3.6286203851e-02 & 0.0000000000e+00 \\
0.0018518519 & 0.0064743298 & 3.5106196263e-02 & 2.3637643620e-05 \\ 
0.0074074074 & 0.0251788386 & 3.1863068261e-02 & 8.5448967297e-05 \\ 
0.0166666667 & 0.0541259986 & 2.7304450667e-02 & 1.6345757744e-04 \\ 
0.0296296296 & 0.0904962838 & 2.2304606782e-02 & 2.3443830054e-04 \\ 
0.0462962963 & 0.1311832131 & 1.7570120169e-02 & 2.8338116595e-04 \\ 
0.0666666667 & 0.1732771212 & 1.3509063981e-02 & 3.0609684764e-04 \\ 
0.0907407407 & 0.2143825635 & 1.0255357455e-02 & 3.0629942762e-04 \\ 
0.1185185185 & 0.2527489948 & 7.7654175911e-03 & 2.9105710804e-04 \\ 
0.1500000000 & 0.2872548584 & 5.9143273246e-03 & 2.6737047585e-04
\end{tabular}
\end{ruledtabular} 
\end{table}

\begin{table}[H]
\caption{Love numbers for $n = 2.00$ and $l=3$. Integration of the
  Newtonian Clairaut equation (for $b=0$) returns $k_{\rm el} 
  = 2.439399851849\e{-02}$. This provides evidence that our results
  for the electric-type Love numbers are accurate to nine
  significant digits. We believe that our results for the magnetic-type Love
  numbers are also accurate to nine significant digits.} 
\begin{ruledtabular} 
\begin{tabular}{llll}
$ b$ & $2M/R$ & $k_{\rm el}$ & $k_{\rm mag}$ \\ 
\hline 
0.0000000000 & 0.0000000000 & 2.4393998521e-02 & 0.0000000000e+00 \\
0.0013580247 & 0.0044806121 & 2.3804409363e-02 & 1.1834173013e-05 \\ 
0.0054320988 & 0.0175403971 & 2.2147566853e-02 & 4.3858418141e-05 \\ 
0.0122222222 & 0.0381005822 & 1.9717948676e-02 & 8.7207170874e-05 \\ 
0.0217283951 & 0.0645673693 & 1.6892301431e-02 & 1.3133222716e-04 \\ 
0.0339506173 & 0.0950758472 & 1.4022630411e-02 & 1.6767660885e-04 \\ 
0.0488888889 & 0.1277344111 & 1.1366324623e-02 & 1.9163622562e-04 \\ 
0.0665432099 & 0.1608206440 & 9.0665499267e-03 & 2.0252062134e-04 \\ 
0.0869135802 & 0.1929044035 & 7.1697064326e-03 & 2.0229022040e-04 \\ 
0.1100000000 & 0.2228982181 & 5.6581213050e-03 & 1.9405485402e-04
\end{tabular}
\end{ruledtabular} 
\end{table}

\begin{table}[H]
\caption{Love numbers for $n = 0.50$ and $l=4$. Integration of the
  Newtonian Clairaut equation (for $b=0$) returns $k_{\rm el} 
  = 1.250625809919\e{-01}$. This provides evidence that our results
  for the electric-type Love numbers are accurate to six significant
  digits. We believe that our results for the magnetic-type Love
  numbers are accurate to nine significant digits.} 
\begin{ruledtabular} 
\begin{tabular}{llll}
$ b$ & $2M/R$ & $k_{\rm el}$ & $k_{\rm mag}$ \\ 
\hline 
0.0000000000 & 0.0000000000 & 1.2506232752e-01 & 0.0000000000e+00 \\ 
0.0162962963 & 0.0627859865 & 8.9880099035e-02 & 4.1259713417e-04 \\ 
0.0651851852 & 0.2085406132 & 3.8670819375e-02 & 6.8285490951e-04 \\ 
0.1466666667 & 0.3636165454 & 1.3511973834e-02 & 4.9948942099e-04 \\ 
0.2607407407 & 0.4883414066 & 4.9038796705e-03 & 2.9045885185e-04 \\ 
0.4074074074 & 0.5772867923 & 2.0751263151e-03 & 1.6878093386e-04 \\ 
0.5866666667 & 0.6379537736 & 1.0476762526e-03 & 1.0600793048e-04 \\ 
0.7985185185 & 0.6789539591 & 6.1942259351e-04 & 7.2980383110e-05 \\ 
1.0429629630 & 0.7068171264 & 4.1610976670e-04 & 5.4565032992e-05 \\ 
1.3200000000 & 0.7259502382 & 3.0850496903e-04 & 4.3643093032e-05
\end{tabular}
\end{ruledtabular} 
\end{table}

\begin{table}[H]
\caption{Love numbers for $n = 0.75$ and $l=4$. Integration of the
  Newtonian Clairaut equation (for $b=0$) returns $k_{\rm el} 
  = 8.731859904775\e{-02}$. This provides evidence that our results
  for the electric-type Love numbers are accurate to eight significant
  digits. We believe that our results for the magnetic-type Love
  numbers are accurate to nine significant digits.} 
\begin{ruledtabular} 
\begin{tabular}{llll}
$ b$ & $2M/R$ & $k_{\rm el}$ & $k_{\rm mag}$ \\ 
\hline 
0.0000000000 & 0.0000000000 & 8.7318599147e-02 & 0.0000000000e+00 \\ 
0.0092592593 & 0.0363293144 & 7.1916151340e-02 & 1.9820540551e-04 \\ 
0.0370370370 & 0.1294393332 & 4.2448736557e-02 & 4.5682358629e-04 \\ 
0.0833333333 & 0.2455878442 & 2.0467648174e-02 & 4.7418211732e-04 \\ 
0.1481481481 & 0.3564603549 & 9.2636229108e-03 & 3.5718798449e-04 \\ 
0.2314814815 & 0.4485200887 & 4.3517890992e-03 & 2.4045653422e-04 \\ 
0.3333333333 & 0.5194393410 & 2.2374880023e-03 & 1.6046670110e-04 \\ 
0.4537037037 & 0.5719839827 & 1.2812234382e-03 & 1.1123202446e-04 \\ 
0.5925925926 & 0.6101589410 & 8.1468581955e-04 & 8.1394894614e-05 \\ 
0.7500000000 & 0.6376107260 & 5.6833824831e-04 & 6.2954932476e-05
\end{tabular}
\end{ruledtabular} 
\end{table}

\begin{table}[H]
\caption{Love numbers for $n = 1.00$ and $l=4$. Integration of the
  Newtonian Clairaut equation (for $b=0$) returns $k_{\rm el} 
  = 6.024125532418\e{-02}$. This provides evidence that our results
  for the electric-type Love numbers are accurate to nine significant
  digits. We believe that our results for the magnetic-type Love
  numbers are also accurate to nine significant digits.} 
\begin{ruledtabular} 
\begin{tabular}{llll}
$ b$ & $2M/R$ & $k_{\rm el}$ & $k_{\rm mag}$ \\ 
\hline 
0.0000000000 & 0.0000000000 & 6.0241255395e-02 & 0.0000000000e+00 \\ 
0.0054320988 & 0.0211887760 & 5.3646913671e-02 & 9.0169742307e-05 \\ 
0.0217283951 & 0.0788326459 & 3.8686876721e-02 & 2.5599084572e-04 \\ 
0.0488888889 & 0.1586178173 & 2.3850649236e-02 & 3.4511609518e-04 \\ 
0.0869135802 & 0.2449940757 & 1.3456371026e-02 & 3.3171170926e-04 \\ 
0.1358024691 & 0.3264977638 & 7.3994944976e-03 & 2.6913831658e-04 \\ 
0.1955555556 & 0.3971100356 & 4.1569144571e-03 & 2.0272875503e-04 \\ 
0.2661728395 & 0.4550360296 & 2.4561332329e-03 & 1.4990535283e-04 \\ 
0.3476543210 & 0.5008905693 & 1.5478883581e-03 & 1.1224767948e-04 \\ 
0.4400000000 & 0.5363092473 & 1.0442398816e-03 & 8.6440104267e-05
\end{tabular}
\end{ruledtabular} 
\end{table}

\begin{table}[H]
\caption{Love numbers for $n = 1.25$ and $l=4$. Integration of the
  Newtonian Clairaut equation (for $b=0$) returns $k_{\rm el} 
  = 4.096746123839\e{-02}$. This provides evidence that our results
  for the electric-type Love numbers are accurate to eight significant
  digits. We believe that our results for the magnetic-type Love
  numbers are accurate to nine significant digits.} 
\begin{ruledtabular} 
\begin{tabular}{llll}
$ b$ & $2M/R$ & $k_{\rm el}$ & $k_{\rm mag}$ \\ 
\hline 
0.0000000000 & 0.0000000000 & 4.0967461120e-02 & 0.0000000000e+00 \\ 
0.0037037037 & 0.0140910881 & 3.7831532658e-02 & 4.4448609345e-05 \\ 
0.0148148148 & 0.0535218477 & 3.0102008525e-02 & 1.3969326868e-04 \\ 
0.0333333333 & 0.1109782733 & 2.1224824306e-02 & 2.1676154291e-04 \\ 
0.0592592593 & 0.1774670027 & 1.3778277197e-02 & 2.4208766237e-04 \\ 
0.0925925926 & 0.2449789236 & 8.5688261330e-03 & 2.2508402313e-04 \\ 
0.1333333333 & 0.3079051472 & 5.2857027758e-03 & 1.8910615297e-04 \\ 
0.1814814815 & 0.3631764781 & 3.3202258257e-03 & 1.5123549667e-04 \\ 
0.2370370370 & 0.4096934360 & 2.1611153793e-03 & 1.1903913573e-04 \\ 
0.3000000000 & 0.4475972736 & 1.4718380690e-03 & 9.4141547749e-05
\end{tabular}
\end{ruledtabular} 
\end{table}

\begin{table}[H]
\caption{Love numbers for $n = 1.50$ and $l=4$. Integration of the
  Newtonian Clairaut equation (for $b=0$) returns $k_{\rm el} 
  = 2.739306738271\e{-02}$. This provides evidence that our results
  for the electric-type Love numbers are accurate to eight significant
  digits. We believe that our results for the magnetic-type Love
  numbers are accurate to nine significant digits.} 
\begin{ruledtabular} 
\begin{tabular}{llll}
$ b$ & $2M/R$ & $k_{\rm el}$ & $k_{\rm mag}$ \\ 
\hline 
0.0000000000 & 0.0000000000 & 2.7393067294e-02 & 0.0000000000e+00 \\ 
0.0025925926 & 0.0095061309 & 2.5904136864e-02 & 2.1590905746e-05 \\ 
0.0103703704 & 0.0366144717 & 2.2022860127e-02 & 7.2661900733e-05 \\ 
0.0233333333 & 0.0775387088 & 1.7085103298e-02 & 1.2456848415e-04 \\ 
0.0414814815 & 0.1272141307 & 1.2356917233e-02 & 1.5600796417e-04 \\ 
0.0648148148 & 0.1805262693 & 8.5360694947e-03 & 1.6255526538e-04 \\ 
0.0933333333 & 0.2332124569 & 5.7656653837e-03 & 1.5121350651e-04 \\ 
0.1270370370 & 0.2822627839 & 3.8858798086e-03 & 1.3142526538e-04 \\ 
0.1659259259 & 0.3259042474 & 2.6550241954e-03 & 1.1010882007e-04 \\ 
0.2100000000 & 0.3633567807 & 1.8598572889e-03 & 9.0858200272e-05
\end{tabular}
\end{ruledtabular} 
\end{table}

\begin{table}[H]
\caption{Love numbers for $n = 1.75$ and $l=4$. Integration of the
  Newtonian Clairaut equation (for $b=0$) returns $k_{\rm el} 
  = 1.795919608352\e{-02}$. This provides evidence that our results
  for the electric-type Love numbers are accurate to eight significant
  digits. We believe that our results for the magnetic-type Love
  numbers are accurate to nine significant digits.} 
\begin{ruledtabular} 
\begin{tabular}{llll}
$ b$ & $2M/R$ & $k_{\rm el}$ & $k_{\rm mag}$ \\ 
\hline 
0.0000000000 & 0.0000000000 & 1.7959195798e-02 & 0.0000000000e+00 \\ 
0.0018518519 & 0.0064743298 & 1.7256414891e-02 & 1.0299492462e-05 \\ 
0.0074074074 & 0.0251788386 & 1.5352372303e-02 & 3.6329038871e-05 \\ 
0.0166666667 & 0.0541259986 & 1.2748788134e-02 & 6.6858496307e-05 \\ 
0.0296296296 & 0.0904962838 & 1.0001909468e-02 & 9.1232683035e-05 \\ 
0.0462962963 & 0.1311832131 & 7.5212893015e-03 & 1.0412456086e-04 \\ 
0.0666666667 & 0.1732771212 & 5.5036193453e-03 & 1.0576629570e-04 \\ 
0.0907407407 & 0.2143825635 & 3.9751203447e-03 & 9.9452555069e-05 \\ 
0.1185185185 & 0.2527489948 & 2.8695308743e-03 & 8.8982268992e-05 \\ 
0.1500000000 & 0.2872548584 & 2.0913685076e-03 & 7.7283983050e-05
\end{tabular}
\end{ruledtabular} 
\end{table}

\begin{table}[H]
\caption{Love numbers for $n = 2.00$ and $l=4$. Integration of the
  Newtonian Clairaut equation (for $b=0$) returns $k_{\rm el} 
  = 1.150774963254\e{-02}$. This provides evidence that our results
  for the electric-type Love numbers are accurate to nine significant
  digits. We believe that our results for the magnetic-type Love
  numbers are also accurate to nine significant digits.} 
\begin{ruledtabular} 
\begin{tabular}{llll}
$ b$ & $2M/R$ & $k_{\rm el}$ & $k_{\rm mag}$ \\ 
\hline 
0.0000000000 & 0.0000000000 & 1.1507749634e-02 & 0.0000000000e+00 \\ 
0.0013580247 & 0.0044806121 & 1.1175986242e-02 & 4.8517346178e-06 \\ 
0.0054320988 & 0.0175403971 & 1.0253199166e-02 & 1.7666877492e-05 \\ 
0.0122222222 & 0.0381005822 & 8.9267030374e-03 & 3.4155212233e-05 \\ 
0.0217283951 & 0.0645673693 & 7.4272319821e-03 & 4.9576781437e-05 \\ 
0.0339506173 & 0.0950758472 & 5.9573506314e-03 & 6.0605384193e-05 \\ 
0.0488888889 & 0.1277344111 & 4.6508432777e-03 & 6.6035419880e-05 \\ 
0.0665432099 & 0.1608206440 & 3.5682657166e-03 & 6.6396107161e-05 \\ 
0.0869135802 & 0.1929044035 & 2.7150400670e-03 & 6.3100849384e-05 \\ 
0.1100000000 & 0.2228982181 & 2.0653404463e-03 & 5.7696936776e-05
\end{tabular}
\end{ruledtabular} 
\end{table}

\begin{table}[H]
\caption{Love numbers for $n = 0.50$ and $l=5$. Integration of the
  Newtonian Clairaut equation (for $b=0$) returns $k_{\rm el} 
  = 8.758378097872\e{-02}$. This provides evidence that our results
  for the electric-type Love numbers are accurate to five significant
  digits. We believe that our results for the magnetic-type Love
  numbers are accurate to nine significant digits.} 
\begin{ruledtabular} 
\begin{tabular}{llll}
$ b$ & $2M/R$ & $k_{\rm el}$ & $k_{\rm mag}$ \\ 
\hline 
0.0000000000 & 0.0000000000 & 8.7583597477e-02 & 0.0000000000e+00 \\ 
0.0162962963 & 0.0627859865 & 5.8953726923e-02 & 2.4566625412e-04 \\ 
0.0651851852 & 0.2085406132 & 2.1502135233e-02 & 3.4017649539e-04 \\ 
0.1466666667 & 0.3636165454 & 6.1438638016e-03 & 2.0040145035e-04 \\ 
0.2607407407 & 0.4883414066 & 1.8484475878e-03 & 9.5309205703e-05 \\ 
0.4074074074 & 0.5772867923 & 6.7146633825e-04 & 4.7055290188e-05 \\ 
0.5866666667 & 0.6379537736 & 3.0201770900e-04 & 2.6143552436e-05 \\ 
0.7985185185 & 0.6789539591 & 1.6415053975e-04 & 1.6470651693e-05 \\ 
1.0429629630 & 0.7068171264 & 1.0383693027e-04 & 1.1563913004e-05 \\ 
1.3200000000 & 0.7259502382 & 7.3777009618e-05 & 8.8490411630e-06
\end{tabular}
\end{ruledtabular} 
\end{table}

\begin{table}[H]
\caption{Love numbers for $n = 0.75$ and $l=5$. Integration of the
  Newtonian Clairaut equation (for $b=0$) returns $k_{\rm el} 
  = 5.904211079675\e{-02}$. This provides evidence that our results
  for the electric-type Love numbers are accurate to nine significant
  digits. We believe that our results for the magnetic-type Love
  numbers are also accurate to nine significant digits.} 
\begin{ruledtabular} 
\begin{tabular}{llll}
$ b$ & $2M/R$ & $k_{\rm el}$ & $k_{\rm mag}$ \\ 
\hline 
0.0000000000 & 0.0000000000 & 5.9042110830e-02 & 0.0000000000e+00 \\ 
0.0092592593 & 0.0363293144 & 4.6830270753e-02 & 1.1653809273e-04 \\ 
0.0370370370 & 0.1294393332 & 2.4976158762e-02 & 2.4051097139e-04 \\ 
0.0833333333 & 0.2455878442 & 1.0491664768e-02 & 2.1488760292e-04 \\ 
0.1481481481 & 0.3564603549 & 4.1030598266e-03 & 1.3812306450e-04 \\ 
0.2314814815 & 0.4485200887 & 1.6845383073e-03 & 8.0340315047e-05 \\ 
0.3333333333 & 0.5194393410 & 7.7279612957e-04 & 4.7382520827e-05 \\ 
0.4537037037 & 0.5719839827 & 4.0394347591e-04 & 2.9755158176e-05 \\ 
0.5925925926 & 0.6101589410 & 2.3943047563e-04 & 2.0179431020e-05 \\ 
0.7500000000 & 0.6376107260 & 1.5846629836e-04 & 1.4743433223e-05
\end{tabular}
\end{ruledtabular} 
\end{table}

\begin{table}[H]
\caption{Love numbers for $n = 1.00$ and $l=5$. Integration of the
  Newtonian Clairaut equation (for $b=0$) returns $k_{\rm el} 
  = 3.929250022713\e{-02}$. This provides evidence that our results
  for the electric-type Love numbers are accurate to nine significant
  digits. We believe that our results for the magnetic-type Love
  numbers are also accurate to nine significant digits.} 
\begin{ruledtabular} 
\begin{tabular}{llll}
$ b$ & $2M/R$ & $k_{\rm el}$ & $k_{\rm mag}$ \\ 
\hline 
0.0000000000 & 0.0000000000 & 3.9292500283e-02 & 0.0000000000e+00 \\ 
0.0054320988 & 0.0211887760 & 3.4232186798e-02 & 5.1595279179e-05 \\ 
0.0217283951 & 0.0788326459 & 2.3214946910e-02 & 1.3691347868e-04 \\ 
0.0488888889 & 0.1586178173 & 1.3083821677e-02 & 1.6725789364e-04 \\ 
0.0869135802 & 0.2449940757 & 6.6517658359e-03 & 1.4339451212e-04 \\ 
0.1358024691 & 0.3264977638 & 3.2897043734e-03 & 1.0355791045e-04 \\ 
0.1955555556 & 0.3971100356 & 1.6738586645e-03 & 6.9962844411e-05 \\ 
0.2661728395 & 0.4550360296 & 9.0663862820e-04 & 4.7012138969e-05 \\ 
0.3476543210 & 0.5008905693 & 5.3118742797e-04 & 3.2481406675e-05 \\ 
0.4400000000 & 0.5363092473 & 3.3782253209e-04 & 2.3432821346e-05
\end{tabular}
\end{ruledtabular} 
\end{table}

\begin{table}[H]
\caption{Love numbers for $n = 1.25$ and $l=5$. Integration of the
  Newtonian Clairaut equation (for $b=0$) returns $k_{\rm el} 
  = 2.574776897544\e{-02}$. This provides evidence that our results
  for the electric-type Love numbers are accurate to nine significant
  digits. We believe that our results for the magnetic-type Love
  numbers are also accurate to nine significant digits.} 
\begin{ruledtabular} 
\begin{tabular}{llll}
$ b$ & $2M/R$ & $k_{\rm el}$ & $k_{\rm mag}$ \\ 
\hline 
0.0000000000 & 0.0000000000 & 2.5747768892e-02 & 0.0000000000e+00 \\ 
0.0037037037 & 0.0140910881 & 2.3431732289e-02 & 2.4484594391e-05 \\ 
0.0148148148 & 0.0535218477 & 1.7882343045e-02 & 7.3469913671e-05 \\ 
0.0333333333 & 0.1109782733 & 1.1838067300e-02 & 1.0630025814e-04 \\ 
0.0592592593 & 0.1774670027 & 7.1175929856e-03 & 1.0903958368e-04 \\ 
0.0925925926 & 0.2449789236 & 4.0764785256e-03 & 9.2520830998e-05 \\ 
0.1333333333 & 0.3079051472 & 2.3180122367e-03 & 7.1006099307e-05 \\ 
0.1814814815 & 0.3631764781 & 1.3500205542e-03 & 5.2196651860e-05 \\ 
0.2370370370 & 0.4096934360 & 8.2182992059e-04 & 3.8121417411e-05 \\ 
0.3000000000 & 0.4475972736 & 5.2873946327e-04 & 2.8280143688e-05
\end{tabular}
\end{ruledtabular} 
\end{table}

\begin{table}[H]
\caption{Love numbers for $n = 1.50$ and $l=5$. Integration of the
  Newtonian Clairaut equation (for $b=0$) returns $k_{\rm el} 
  = 1.656876321404\e{-02}$. This provides evidence that our results
  for the electric-type Love numbers are accurate to eight significant
  digits. We believe that our results for the magnetic-type Love
  numbers are accurate to nine significant digits.} 
\begin{ruledtabular} 
\begin{tabular}{llll}
$ b$ & $2M/R$ & $k_{\rm el}$ & $k_{\rm mag}$ \\ 
\hline 
0.0000000000 & 0.0000000000 & 1.6568763135e-02 & 0.0000000000e+00 \\ 
0.0025925926 & 0.0095061309 & 1.5513747923e-02 & 1.1396769321e-05 \\ 
0.0103703704 & 0.0366144717 & 1.2817102827e-02 & 3.7144252136e-05 \\ 
0.0233333333 & 0.0775387088 & 9.5123354123e-03 & 6.0593529815e-05 \\ 
0.0414814815 & 0.1272141307 & 6.5071687257e-03 & 7.1292599566e-05 \\ 
0.0648148148 & 0.1805262693 & 4.2236890233e-03 & 6.9268626197e-05 \\ 
0.0933333333 & 0.2332124569 & 2.6751245641e-03 & 5.9938429831e-05 \\ 
0.1270370370 & 0.2822627839 & 1.6935347048e-03 & 4.8543637402e-05 \\ 
0.1659259259 & 0.3259042474 & 1.0918134805e-03 & 3.8082191743e-05 \\ 
0.2100000000 & 0.3633567807 & 7.2625290353e-04 & 2.9627611402e-05
\end{tabular}
\end{ruledtabular} 
\end{table}

\begin{table}[H]
\caption{Love numbers for $n = 1.75$ and $l=5$. Integration of the
  Newtonian Clairaut equation (for $b=0$) returns $k_{\rm el} 
  = 1.043995446810\e{-02}$. This provides evidence that our results
  for the electric-type Love numbers are accurate to nine significant
  digits. We believe that our results for the magnetic-type Love
  numbers are also accurate to nine significant digits.} 
\begin{ruledtabular} 
\begin{tabular}{llll}
$ b$ & $2M/R$ & $k_{\rm el}$ & $k_{\rm mag}$ \\ 
\hline 
0.0000000000 & 0.0000000000 & 1.0439954387e-02 & 0.0000000000e+00 \\ 
0.0018518519 & 0.0064743298 & 9.9635882364e-03 & 5.1907364491e-06 \\ 
0.0074074074 & 0.0251788386 & 8.6906171854e-03 & 1.7903644602e-05 \\ 
0.0166666667 & 0.0541259986 & 6.9954658202e-03 & 3.1806852682e-05 \\ 
0.0296296296 & 0.0904962838 & 5.2724767209e-03 & 4.1473311159e-05 \\ 
0.0462962963 & 0.1311832131 & 3.7857836403e-03 & 4.4914541070e-05 \\ 
0.0666666667 & 0.1732771212 & 2.6366800282e-03 & 4.3129384841e-05 \\ 
0.0907407407 & 0.2143825635 & 1.8116921836e-03 & 3.8310103373e-05 \\ 
0.1185185185 & 0.2527489948 & 1.2463184219e-03 & 3.2437461311e-05 \\ 
0.1500000000 & 0.2872548584 & 8.6864853073e-04 & 2.6760732768e-05
\end{tabular}
\end{ruledtabular} 
\end{table}

\begin{table}[H]
\caption{Love numbers for $n = 2.00$ and $l=5$. Integration of the
  Newtonian Clairaut equation (for $b=0$) returns $k_{\rm el} 
  = 6.419966834096\e{-03}$. This provides evidence that our results
  for the electric-type Love numbers are accurate to nine significant
  digits. We believe that our results for the magnetic-type Love
  numbers are also accurate to nine significant digits.} 
\begin{ruledtabular} 
\begin{tabular}{llll}
$ b$ & $2M/R$ & $k_{\rm el}$ & $k_{\rm mag}$ \\ 
\hline 
0.0000000000 & 0.0000000000 & 6.4199668350e-03 & 0.0000000000e+00 \\ 
0.0013580247 & 0.0044806121 & 6.2054683813e-03 & 2.3273253180e-06 \\ 
0.0054320988 & 0.0175403971 & 5.6146736603e-03 & 8.3410581918e-06 \\ 
0.0122222222 & 0.0381005822 & 4.7814250834e-03 & 1.5722362491e-05 \\ 
0.0217283951 & 0.0645673693 & 3.8647546556e-03 & 2.2075823575e-05 \\ 
0.0339506173 & 0.0950758472 & 2.9960238568e-03 & 2.5950234182e-05 \\ 
0.0488888889 & 0.1277344111 & 2.2531557240e-03 & 2.7084179213e-05 \\ 
0.0665432099 & 0.1608206440 & 1.6628187985e-03 & 2.6037432070e-05 \\ 
0.0869135802 & 0.1929044035 & 1.2172373651e-03 & 2.3660856250e-05 \\ 
0.1100000000 & 0.2228982181 & 8.9227889200e-04 & 2.0720823774e-05
\end{tabular}
\end{ruledtabular} 
\end{table}

\endgroup 

\bibliography{../bib/master} 

\begin{thebibliography}{10}
\expandafter\ifx\csname url\endcsname\relax
  \def\url#1{{\tt #1}}\fi
\expandafter\ifx\csname urlprefix\endcsname\relax\def\urlprefix{URL }\fi

\bibitem{flanagan-hinderer:08}
E.~E. Flanagan and T.~Hinderer, {\em Constraining neutron star tidal Love
  numbers with gravitational wave detectors\/}, Phys. Rev. D {\bf 77},
  021502(R) (2008), arXiv:0709.1915.

\bibitem{hinderer:08}
T.~Hinderer, {\em Tidal Love numbers of neutron stars\/}, Astrophys. J. {\bf
  677}, 1216 (2008), erratum: Astrophys. J. {\bf 697}, 964 (2009),
  arXiv:0711.2420.

\bibitem{love:11}
A.~E.~H. Love, {\em Some problems of geodynamics\/} (Cornell University
  Library, Ithaca, USA, 1911).

\bibitem{murray-dermott:99}
C.~D. Murray and S.~F. Dermott, {\em Solar System Dynamics\/} (Cambridge
  University Press, Cambridge, England, 1999).

\bibitem{damour-soffel-xu:92}
T.~Damour, M.~Soffel, and C.~Xu, {\em General-relativistic celestial mechanics.
  II. Translational equations of motion\/}, Phys. Rev. D {\bf 45}, 1017 (1992).

\bibitem{favata:06}
M.~Favata, {\em Are neutron stars crushed? Gravitomagnetic tidal fields as a
  mechanism for binary-induced collapse\/}, Phys. Rev. D {\bf 73}, 104005
  (2006), arXiv:astro-ph/0510668.

\bibitem{poisson:05}
E.~Poisson, {\em Metric of a tidally distorted, nonrotating black hole\/},
  Phys. Rev. Lett. {\bf 94}, 161103 (2005), arXiv:gr-qc/0501032.

\bibitem{preston-poisson:06b}
B.~Preston and E.~Poisson, {\em A light-cone gauge for black-hole perturbation
  theory\/}, Phys. Rev. D {\bf 74}, 064010 (2006), arXiv:gr-qc/0606094.

\bibitem{zhang:86}
X.-H. Zhang, {\em Multipole expansions of the general-relativistic
  gravitational field of the external universe\/}, Phys. Rev. D {\bf 34}, 991
  (1986).

\bibitem{thorne-campolattaro:67}
K.~S. Thorne and A.~Campolattaro, {\em Non-radial pulsation of general
  relativistic stellar models. I. Analytical analysis for $l \geq 2$\/},
  Astrophys. J. {\bf 149}, 591 (1967).

\bibitem{brooker-olle:55}
R.~A. Brooker and T.~W. Olle, {\em Apsidal-motion constants for polytropic
  models\/}, Mon. Not. Roy. Astron. Soc. {\bf 115}, 101 (1955).

\bibitem{damour-nagar:09}
T.~Damour and A.~Nagar, {\em Relativistic tidal properties of neutron stars\/}
  (2009), arXiv:0906.0096.

\bibitem{fang-lovelace:05}
H.~Fang and G.~Lovelace, {\em Tidal coupling of a Schwarzschild black hole and
  circularly orbiting moon\/}, Phys. Rev. D {\bf 72}, 124016 (2005),
  arXiv:gr-qc/0505156.

\bibitem{thorne-hartle:85}
K.~S. Thorne and J.~B. Hartle, {\em Laws of motion and precession for black
  holes and other bodies\/}, Phys. Rev. D {\bf 31}, 1815 (1985).

\bibitem{zhang:85}
X.-H. Zhang, {\em Higher-order corrections to the laws of motion and precession
  for black holes and other bodies\/}, Phys. Rev. D {\bf 31}, 3130 (1985).

\bibitem{suen:86b}
W.-M. Suen, {\em Distorted black holes in terms of multipole moments\/}, Phys.
  Rev. D {\bf 34}, 3633 (1986).

\bibitem{martel-poisson:05}
K.~Martel and E.~Poisson, {\em Gravitational perturbations of the Schwarzschild
  spacetime: A practical covariant and gauge-invariant formalism\/}, Phys. Rev.
  D {\bf 71}, 104003 (2005), arXiv:gr-qc/0502028.

\bibitem{grtensor}
GrTensorII, developed by Peter Musgrave, Denis Pollney and Kayll Lake, is
  available free of charge at http://grtensor.org/.

\bibitem{numerical-recipes:92}
W.~H. Press, S.~A. Teukolsky, W.~T. Vetterling, and B.~P. Flannery, {\em
  Numerical recipes in C++: The Art of Scientific Computing\/} (Cambridge
  University Press, Cambridge, England, 2002).

\bibitem{GSL}
The GNU Scientific Library is a numerical library for C and C++ programmers. It
  is free software under the GNU General Public License. The library is
  available for download at http://www.gnu.org/software/gsl/.

\end{thebibliography}
\end{document}